\documentclass[apj]{emulateapj}
\usepackage{color,graphicx,textcomp}
\newcommand{\bjdtdb}{\ensuremath{\rm {BJD_{TDB}}}}

\newcommand{\ms}{m s$^{-1}$}

\begin{document}


\title{\textit{Spitzer} Secondary Eclipse Observations of Five Cool Gas Giant Planets and Empirical Trends in Cool Planet Emission Spectra}

\author{
Joshua~A.~Kammer\altaffilmark{1,2}, 
Heather~A.~Knutson\altaffilmark{1,*},
Michael~R.~Line\altaffilmark{3},
Jonathan~J.~Fortney\altaffilmark{3},
Drake~Deming\altaffilmark{4},
Adam~Burrows\altaffilmark{5},
Nicolas~B.~Cowan\altaffilmark{6},
Amaury~H.~M.~J.~Triaud\altaffilmark{7,8}
Eric~Agol\altaffilmark{9},
Jean-Michel~Desert\altaffilmark{10},
Benjamin~J.~Fulton\altaffilmark{11},
Andrew~W.~Howard\altaffilmark{11},
Gregory~P.~Laughlin\altaffilmark{3},
Nikole~K.~Lewis\altaffilmark{12},
Caroline~V.~Morley\altaffilmark{3},
Julianne~I.~Moses\altaffilmark{13},
Adam~P.~Showman\altaffilmark{14},
Kamen~O.~Todorov\altaffilmark{15}
}

\altaffiltext{1}{Division of Geological and Planetary Sciences, California Institute of Technology, Pasadena, CA 91125, USA}
\altaffiltext{2}{Space Studies Department, Southwest Research Institute, Boulder, CO 80302, USA}
\altaffiltext{3}{Department of Astronomy and Astrophysics, University of California at Santa Cruz, Santa Cruz, CA 95064, USA}
\altaffiltext{4}{Department of Astronomy, University of Maryland at College Park, College Park, MD 20742, USA}
\altaffiltext{5}{Department of Astrophysical Sciences, Princeton University, Princeton, NJ 08544, USA}
\altaffiltext{6}{Department of Physics and Astronomy, Amherst College, Amherst, MA 01002, USA}
\altaffiltext{7}{Centre for Planetary Sciences, University of Toronto at Scarborough, 1265 Military Trail, Toronto, ON, M1C 1A4, Canada}
\altaffiltext{8}{Department of Astronomy \& Astrophysics, University of Toronto, Toronto, Ontario M5S 3H4, Canada}
\altaffiltext{9}{Department of Astronomy, University of Washington, Seattle, WA 98195, USA}
\altaffiltext{10}{CASA, Department of Astrophysical and Planetary Sciences, University of Colorado, Boulder, CO 80309, USA}
\altaffiltext{11}{Institute for Astronomy, University of Hawaii, Honolulu, HI 96822, USA}
\altaffiltext{12}{Space Telescope Science Institute, Baltimore, MD 21218, USA}
\altaffiltext{13}{Space Science Institute, Boulder, CO 80301, USA}
\altaffiltext{14}{Department of Planetary Sciences and Lunar and Planetary Laboratory, University of Arizona, Tucson, AZ 85721, USA}
\altaffiltext{15}{Institute for Astronomy, ETH Z\"{u}rich, 8093 Z\"{u}rich, Switzerland}
\altaffiltext{*}{Corresponding author, hknutson@caltech.edu}

\begin{abstract}

In this work we present \textit{Spitzer} 3.6 and 4.5~\micron~secondary eclipse observations of five new cool ($<1200$ K) transiting gas giant planets: \mbox{HAT-P-19b}, \mbox{WASP-6b}, \mbox{WASP-10b}, \mbox{WASP-39b}, and \mbox{WASP-67b}.  We compare our measured eclipse depths to the predictions of a suite of atmosphere models and to eclipse depths for planets with previously published observations in order to constrain the temperature- and mass-dependent properties of gas giant planet atmospheres.   We find that the dayside emission spectra of planets less massive than Jupiter require models with efficient circulation of energy to the night side and/or increased albedos, while those with masses greater than that of Jupiter are consistently best-matched by models with inefficient circulation and low albedos.  At these relatively low temperatures we expect the atmospheric CH$_4$/CO ratio to vary as a function of metallicity, and we therefore use our observations of these planets to constrain their atmospheric metallicities.  We find that the most massive planets have dayside emission spectra that are best-matched by solar metallicity atmosphere models, but we are not able to place strong constraints on metallicities of the smaller planets in our sample.  Interestingly, we find that the ratio of the 3.6 and 4.5~\micron~brightness temperatures for these cool transiting planets is independent of planet temperature, and instead exhibits a tentative correlation with planet mass.  If this trend can be confirmed, it would suggest that the shape of these planets' emission spectra depends primarily on their masses, consistent with the hypothesis that lower-mass planets are more likely to have metal-rich atmospheres.  

\end{abstract}

\keywords{eclipses - planetary systems - techniques: photometric}

\section{Introduction}

\begin{deluxetable*}{lccccccc}
\tabletypesize{\scriptsize}
\tablecaption{Target System Properties\label{table:tab1}}
\tablecolumns{8}
\tablewidth{0pt}
\tablehead{
\colhead{Target} & \colhead{Stellar T$_{eff}$ (K)} & \colhead{$M_p$ (M$_{\rm Jup})$} & \colhead{Period (days)} & \colhead{$T_{c}$\tablenotemark{a}} & \colhead{\textit{e}} & \colhead{T$_{eq}$ (K)\tablenotemark{b}} & \colhead{Ref.} }
\startdata
HAT-P-19b & $5007\pm66$ & $0.292\pm0.018$ & $4.008778\pm6\times10^{-6}$ & $5091.5342\pm0.0003$ & $0.067\pm0.042$ & 1010 & \tablenotemark{c}\\
WASP-6b & $5375\pm65$ & $0.485\pm0.027$ & $3.36100208\pm31\times10^{-8}$ & $4425.02180\pm0.00011$ & $0.041\pm0.019$ & 1184 & \tablenotemark{d}\\
WASP-10b & $4735\pm69$ & $3.14\pm0.27$ & $3.09272932\pm32\times10^{-8}$ & $4664.038090\pm0.000048$ & $0.0473^{+0.0034}_{-0.0029}$ & 980 & \tablenotemark{e}\\
WASP-39b & $5400\pm150$ & $0.284\pm0.031$ & $4.0552965\pm10\times10^{-6}$ & $5342.96956\pm0.00020$ & 0 (fixed) & 1116 & \tablenotemark{f}\\
WASP-67b & $5200\pm100$ & $0.406\pm0.035$ & $4.6144109\pm27\times10^{-7}$ & $5824.37424\pm0.00022$ & 0 (fixed) & 1004 & \tablenotemark{g}
\enddata
\tablenotetext{a}{Measured time of transit center, BJD$_{\rm UTC}$ - 2,450,000.  The reported times for WASP-6b, WASP-10b, and WASP-67b are given in BJD$_{\rm TDB}$, and can be converted to the UTC time at the epoch of our observations by subtracting 66.2~s, XX~s, and XX~s, respectively.}
\tablenotetext{b}{Planet dayside equilibrium temperature calculated assuming zero albedo and uniform heat redistribution.}
\tablenotetext{c}{\cite{hartman2011hat,torres12}}
\tablenotetext{d}{\cite{gillon2009discovery,husnoo2012observational,doyle13,tregloan15}}
\tablenotetext{e}{\cite{torres12,barros13,knutson2014b}}
\tablenotetext{f}{\cite{faedi2011wasp,ricci15}}
\tablenotetext{g}{\cite{hellier12,mancini14}}
\end{deluxetable*}

Observations of the thermal emission spectra of gas giant planets provide an invaluable tool for probing their atmospheric compositions and pressure-temperature profiles.  To date nearly one hundred transiting planets have been observed in secondary eclipse with the \textit{Spitzer Space Telescope}, but the vast majority of these objects have been Jovian-mass planets with temperatures between 1500-2500 K  \citep{madhusudhan2014protostars}.  The atmospheres of smaller and cooler planets remain mostly uncharted territory, in part because it is extremely challenging to detect their thermal emission at the near-infrared wavelengths accessible to most telescopes.  Equilibrium chemistry models predict that between 1000-1200 K the dominant reservoir of atmospheric carbon shifts from CO to CH$_4$, a phenomenon similar to that which occurs in the atmospheres of brown dwarfs \citep{kirkpatrick2005new}.  However, these predictions are dependent not only on temperature but also on the underlying elemental abundances, and therefore are sensitive to changes in atmospheric metallicity.  For cool planets where equilibrium chemistry would normally predict a methane-dominated atmosphere, a decrease in the relative hydrogen abundance will result in reduced CH$_4$ and enhanced CO abundances relative to the solar metallicity predictions \citep{moses2013compositional,hu14,agundez14,venot14}.

This hypothesis was first proposed to explain the emission spectrum of the Neptune-mass planet GJ~436b, which appears to have a CO-rich and methane-poor atmosphere despite its relatively low (670~K) predicted dayside equilibrium temperature \citep{stevenson2010possible,madhusudhan2011}.  Although early studies suggested that disequilibrium chemistry might be responsible, \citet{line2011thermo} demonstrated that these processes alone were insufficient to explain the observed abundances for atmospheric metallicities up to $50\times$ solar.  \cite{moses2013compositional} explored atmosphere models spanning a broad range of metallicities and found that those with metallicities of $300\times$ solar or higher could produce a good match to GJ 436b's dayside emission spectrum.   Although a recent global re-analysis of all available \textit{Spitzer} photometry for GJ 436b by \citet{lanotte2014} found a smaller eclipse depth at 3.6~\micron~than Stevenson et al., this study concludes that the revised emission spectrum is still best described by an atmosphere model with very little methane and significant CO.  To date secondary eclipse detections have been published for three additional gas giant planets with predicted equilibrium temperatures cooler than 1200 K, including WASP-8b \citep{cubillos2013wasp}, WASP-80b \citep{triaud15}, and HAT-P-20b \citep{deming2014spitzer}, but none of these planets appear to share GJ~436b's unique atmospheric chemistry.  Although HAT-P-12b was also observed with \textit{Spitzer}, the secondary eclipse was not detected in either the 3.6 or 4.5~\micron~bands \citep{todorov2013warm}. 

Studies of hydrogen-dominated atmospheres in our own solar system have shown that as core mass fraction increases, so does the atmospheric metallicity.  Uranus and Neptune have atmospheres with carbon to hydrogen ratios approximately $70-100$ times that of the solar value, while Jupiter's atmospheric carbon to hydrogen ratio is only four times that of the sun \citep[e.g.,][]{wong04,fletcher09,karkoschka11,sromovsky11}.  Although we have relatively few direct constraints on the atmospheric metallicities of extrasolar gas giant planets, current mass and radius measurements suggest an inverse relationship between the planet's bulk metallicity and its mass \citep{miller2011}.  For sub-Neptune-sized planets, results from the Kepler telescope indicate that average densities continue to increase with decreasing mass, with the caveat that small planets also appear to exhibit a greater diversity of densities at a fixed mass than their larger gas giant counterparts \citep[e.g.][]{marcy2014,weiss2014,lopez2014,hadden2014,wolfgang2014}.  In this scenario we would not expect WASP-8b, WASP-80b, or HAT-P-20b to share Neptune-mass GJ 436b's unique dayside chemistry, as these three planets all have masses greater than 0.5 M$_{\rm Jup}$.   

It currently remains an open question as to whether or not the metallicities of exoplanet atmospheres increase with decreasing mass, as models of planet formation indicate that small variations in protoplanetary disks can lead to a broad range of planet masses and compositions \citep{fortney13,helled2014formation}.  Measurements of average density alone are not sufficient to uniquely determine atmospheric metallicity, as there is a degeneracy with the assumed composition and mass of the planet's metal-rich core \citep[e.g.,][]{rogers2010,figueria2009,nettelmann2010,valencia2013,howe14}.  Fortunately the combination of transmission spectroscopy and secondary eclipse observations provides unique leverage to directly constrain the atmospheric metallicities of transiting planets \citep[e.g.,][]{kreidberg2014}.  For planets with diffuse, high altitude cloud layers, secondary eclipse spectroscopy offers an added advantage as these cloud layers are less likely to be optically thick when viewed face-on than during transit \citep{fortney2005}.  Similarly, planets with high mean molecular weight atmospheres may have absorption features that are undetectable in transit with current telescopes.  The Neptune-mass planet GJ 436b serves as a useful illustration of this concept, as it has a featureless transmission spectrum but exhibits strong molecular absorption features in its dayside emission spectrum \citep{stevenson2010possible,knutson2014}.

\begin{figure}[h]
\epsscale{1.2}
\plotone{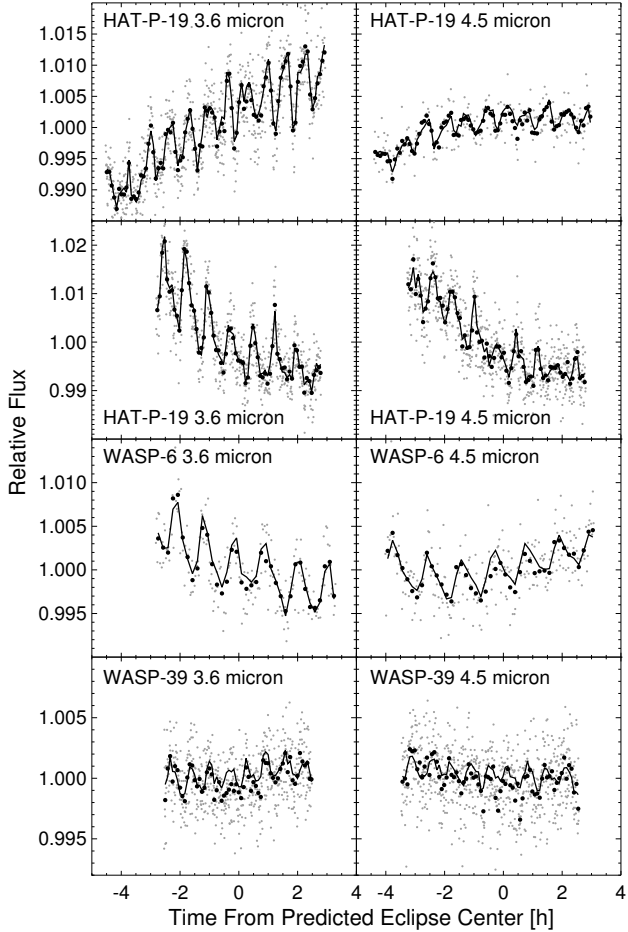}
\caption{Raw \textit{Spitzer} 3.6 and 4.5 $\mu$m photometry as a function of time from the predicted center of eclipse for HAT-P-19b, WASP-6b, and WASP-39b. The relative flux from each star is shown binned in 30~s (grey filled circles) and 5 minute (black filled circles) intervals, and the best fit instrumental models are binned in 5 minute intervals and overplotted for comparison (solid lines).  Intrapixel sensitivity variations cause distinct sawtooth patterns as a result of oscillations in the centroid position of the star with a period of approximately 45 minutes.  Observations for HAT-P-19 are plotted in chronological order, with visits from 2011-2012 one row up from the bottom and visits from 2014 in the bottom row.}
\label{rawfig}
\end{figure}

\begin{figure}[h]
\epsscale{1.2}
\plotone{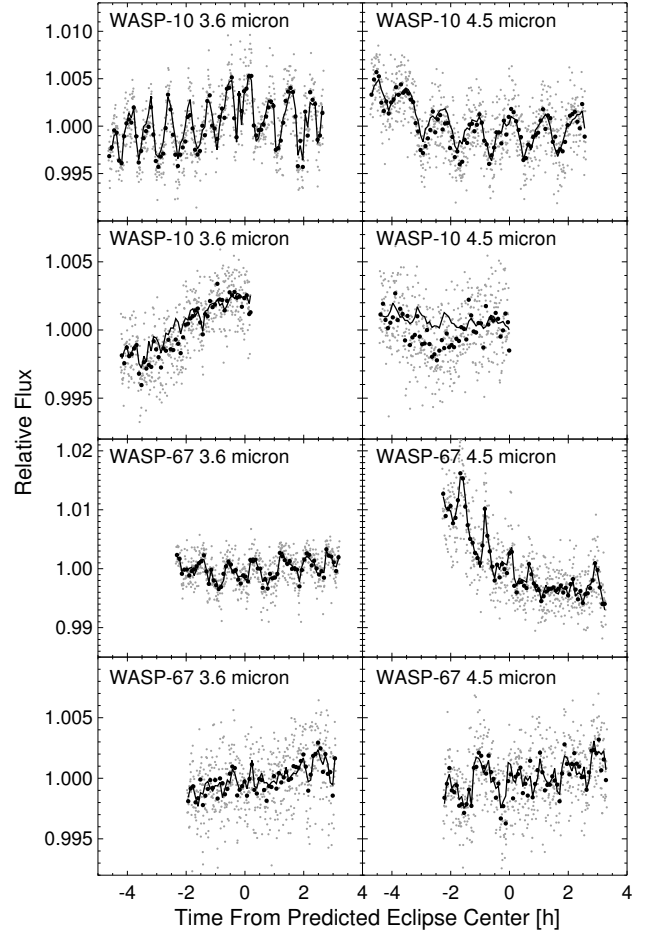}
\caption{Raw \textit{Spitzer} 3.6 and 4.5 $\mu$m photometry as a function of time from the predicted center of eclipse for WASP-10b and WASP-67b; see Fig. \ref{rawfig} for more details.  Duplicate observations for both planets are plotted in chronological order, with earlier visits in the upper row and later visits in the lower row.}
\label{rawfig2}
\end{figure}

\begin{figure}[h]
\epsscale{1.2}
\plotone{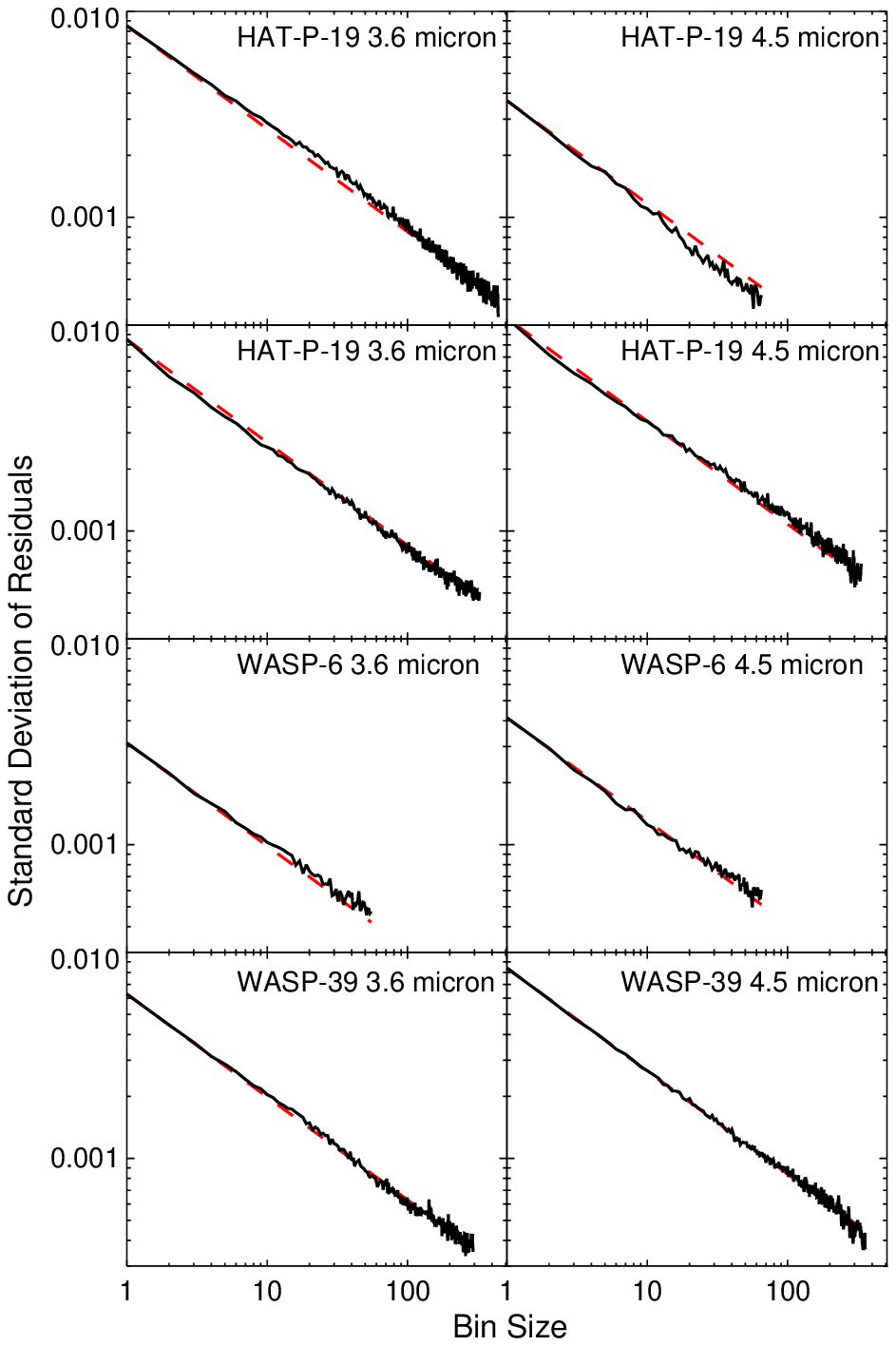}
\caption{Standard deviation of the residuals for HAT-P-19, WASP-6, and WASP-39 after the best-fit eclipse and instrumental noise models have been subtracted as a function of bin size (black lines).  We also show the expected $1/\sqrt n$ scaling for Gaussian measurement errors and no correlation between adjacent photometric points (white noise) as red dashed lines, where we have normalized both curves to match the standard deviation of the unbinned residuals.  Observations for HAT-P-19 are plotted in chronological order, with visits from 2011-2012 one row up from the bottom and visits from 2014 in the bottom row.}
\label{rootn_fig}
\end{figure}

Observations of transiting planets spanning a wide range of effective temperatures and masses can also reveal correlations between the efficiency of atmospheric circulation and other properties of the system.  Tidal evolution models predict that the tidal locking time scales for short period planets should be significantly shorter than the ages of these systems \citep{guillot96,showman15}.  By measuring the day side temperatures of a large, diverse sample of transiting planets, we can place constraints on the nature of the dynamical processes that transport heat between hemispheres, as well as their dependence on the incident stellar flux, planetary mass, metallicity, and other parameters.    Previous studies by \citet{cowan11}, \citet{perez13}, and \citet{schwartz15} indicate that, on average, highly irradiated hot Jupiters appear to have inefficient day-night heat recirculation, whereas more weakly irradiated hot Jupiters seem to exhibit more efficient day-night circulation albeit with a greater degree of diversity.  This trend is in good agreement with results from three-dimensional general circulation models of hot Jupiters including radiative transfer \citep[e.g.,][]{showman09,showman13,showman15,heng12,rauscher12,perna12} and can be explained with simple dynamical theories \citep{perez13}.   Moreover, circulation models with non-grey radiative transfer suggest that both the degree of deviation from radiative equilibrium on the planet's day side and the overall magnitude of the infrared day-night flux difference depend not only on stellar irradiation but also on planetary rotation rate \citep{showman15} and atmospheric metallicity \citep{lewis10}.  Although there are multiple studies \citep{cowan11,perez13,schwartz15} that have explored the observed correlation with incident flux  for relatively hot planets, there are currently very few published observations constraining the efficiency of the day-night circulation on cooler planets as well as those with metal-rich atmospheres.

\begin{figure}[h]
\epsscale{1.2}
\plotone{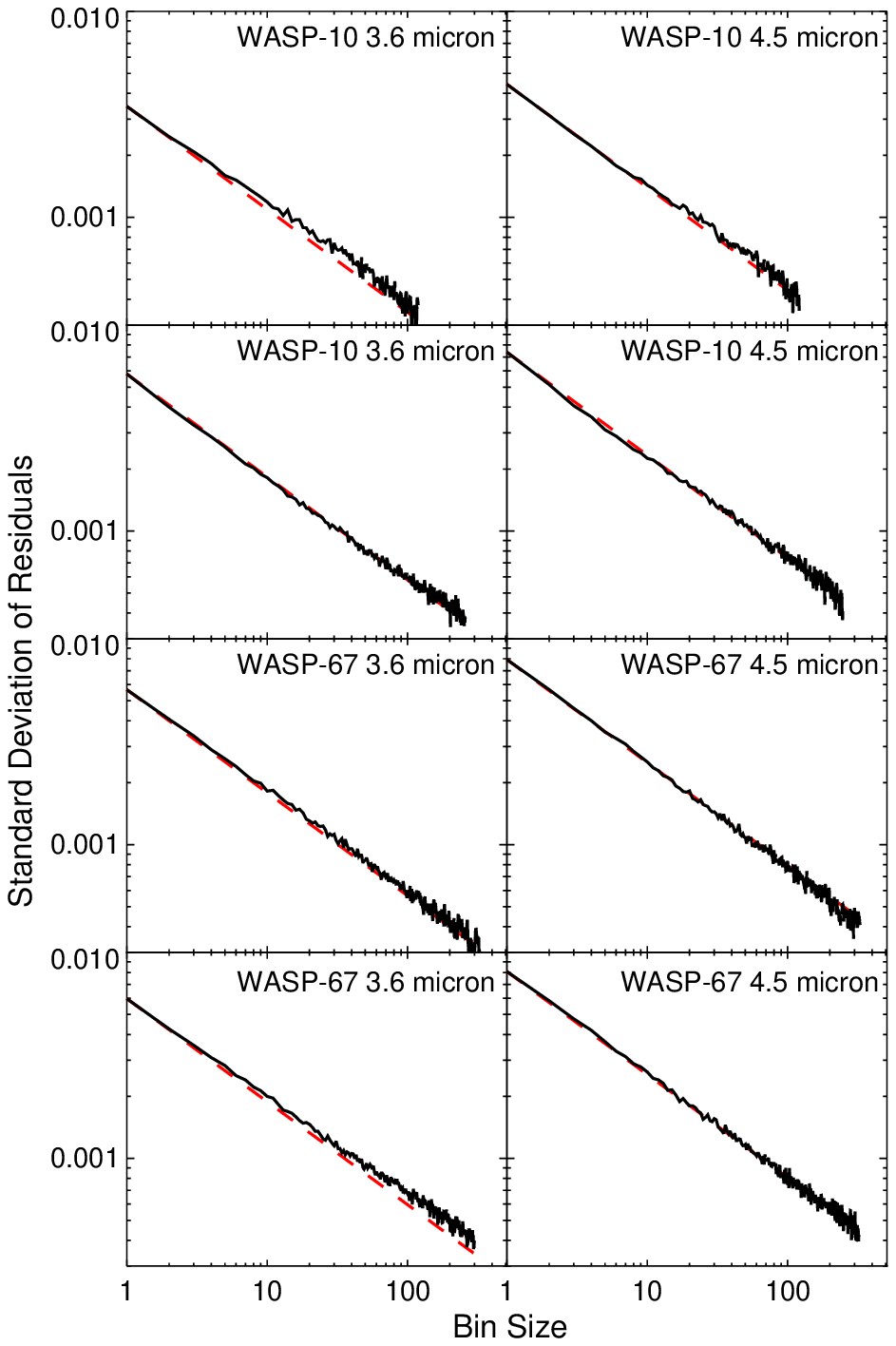}
\caption{Standard deviation of the residuals for WASP-10 and WASP-67 after the best-fit eclipse and instrumental noise models have been subtracted as a function of bin size.  Observations for both planets are plotted in chronological order, with earlier visits in the upper row and later visits in the lower row.  See Fig. \ref{rootn_fig} for more details.}
\label{rootn_fig2}
\end{figure}

In this work we present new \textit{Spitzer} secondary eclipse observations of the cool gas giant planets  \mbox{HAT-P-19b}, \mbox{WASP-6b}, \mbox{WASP-10b}, \mbox{WASP-39b}, and \mbox{WASP-67b}, which have predicted equilibrium temperatures between $900-1200$~K and masses of $0.3-3.1$~M$_{\rm Jup}$.  These observations were obtained as part of a larger program designed to search for correlations between planet mass and atmospheric metallicity (GO 10054, PI Knutson), with a total sample size of fifteen planets.  The selected planets have predicted equilibrium temperatures between 650-1150 K, and if they have hydrogen-rich atmospheres whose chemistry is in local thermodynamic equilibrium we would expect to see strong methane absorption features in their emission spectra.  We list relevant characteristics for the five planets in this study in Table \ref{table:tab1}. In Section \ref{sec:obsdata}, we describe the \textit{Spitzer} data acquisition and reduction, while Sections \ref{sec:xDiscussion} and \ref{sec:xConclusions} discuss the implications of these results for our understanding of the relationship between planet mass and atmospheric metallicity in the gas giant regime.

\begin{deluxetable*}{lcccccccccccc}
\tabletypesize{\scriptsize}
\tablecaption{\textit{Spitzer} Observation Details\label{table:tab2}}
\tablewidth{0pt}
\tablehead{
\colhead{Target} & \colhead{$\lambda$ ($\mu m$)} & \colhead{UT Start Date} & \colhead{Length (h)} & \colhead{$n_{img}$\tablenotemark{a}} & \colhead{$t_{int}$ (s)\tablenotemark{b}} & \colhead{$t_{trim} $\tablenotemark{c}} & \colhead{$n_{bin} $\tablenotemark{c}} & \colhead{$r_{pos}$\tablenotemark{c}} & \colhead{$r_{npix}$\tablenotemark{c}} & \colhead{Fixed\tablenotemark{d}} & \colhead{$r_{phot}$\tablenotemark{c}} & \colhead{Bkd ($\%$)\tablenotemark{e}}}
\startdata
HAT-P-19b & 3.6 & 2012-03-12 & 7.9 & 14144 & 2.0 & 0.5 & 128 & 2.5 & none & yes & 2.2 & 1.88\\
& 3.6 & 2014-10-02 & 6.1 & 10816 & 2.0 & 0.5 & 16 & 1.5 & 2.5 & no & 1.4 & 0.91\\
 & 4.5 & 2011-09-10 & 7.9 & 2167 & 12.0 & 0.5 & 8 & 2.5 & 2.5 & no & 2.5 & 0.82\\
 & 4.5 & 2014-09-28 & 6.1 & 10816 & 2.0 & 0.0 & 32 & 1.5 & 2.5 & no & 1.4 & 0.65\\
WASP-6b & 3.6 & 2009-12-30 & 7.6 & 2096 & 12.0 & 1.5 & 8 & 1.5 & none & yes & 2.3 & $-0.38$\\
 & 4.5 & 2010-01-23 & 7.6 & 2096 & 12.0 & 0.5 & 5 & 1.5 & none & yes & 2.1 & 1.47\\
 WASP-10b & 3.6 & 2011-01-19 & 7.8 & 3976 & 6.0 & 0.5 & 64 & 1.5 & none & yes & 2.1 & -0.17\\
& 3.6 & 2014-09-10 & 4.9 & 8704 & 2.0 & 0.5 & 64 & 1.5 & 1.9 & no & 2.2& 1.47\\
 & 4.5 & 2010-09-17 & 7.8 & 3976 & 6.0 & 0.5 & 64 & 1.5 & none & yes & 2.7 & -0.18\\
 & 4.5 & 2014-09-22 & 4.9 & 8704 & 2.0 & 0.5 & 4 & 1.5 & none & yes & 2.3& 0.39\\
WASP-39b & 3.6 & 2012-04-04 & 7.0 & 12480 & 2.0 & 2.0 & 32 & 2.5 & 3.0 & no & 1.8 & 1.29\\
 & 4.5 & 2012-04-08  & 7.0 & 12480 & 2.0 & 1.0 & 64 & 1.5 & none & yes & 2.4 & 1.20\\
 WASP-67b & 3.6 & 2014-06-28 & 6.0 & 10752 & 2.0 & 0.5 & 8 & 1.5 & none & yes & 2.2 & 1.49\\
& 3.6 & 2014-07-21 & 6.0 & 10752 & 2.0 & 1.0 & 8 & 1.5 & none & yes & 2.3 & 2.10\\
 & 4.5 & 2014-07-07 & 6.0 & 10752 & 2.0 & 0.5 & 16 & 2.0 & 2.1 & no & 1.3 & 0.63\\
 & 4.5 & 2014-07-30 & 6.0 & 10752 & 2.0 & 0.5 & 64 & 2.0 & none & yes & 3.0 & 0.97
\enddata
\tablenotetext{a}{Total number of images.}
\tablenotetext{b}{Image integration time.}
\tablenotetext{c}{$t_{trim}$ is the amount of time in hours trimmed from the start of each time series, $n_{bin}$ is the bin size used in the photometric fits, $r_{pos}$ is the radius of the aperture used to determine the position of the star on the array, $r_{npix}$ is the radius of the aperture used to calculate the noise pixel parameter, and $r_{phot}$ is the radius of the photometric aperture (we provide the median aperture radius over the observation for the time-varying aperture case).  All radii are given in pixels.}
\tablenotetext{d}{Denotes whether the photometry was obtained using a fixed or time-varying aperture.}
\tablenotetext{e}{Sky background contribution to the total flux measured in the selected aperture.}
\end{deluxetable*}

\section{Observations and Data Analysis}\label{sec:obsdata}

\subsection{Photometry and Instrumental Model}

These observations were obtained in the 3.6 and 4.5 $\mu$m bandpasses using the Infra-Red Array Camera (IRAC) on the \textit{Spitzer Space Telescopes}.  HAT-P-19 was observed in full array mode with no peak-up for the 2011 4.5~\micron~eclipse, and in subarray mode with peak-up pointing (this pointing adjustment corrects the initial telescope pointing in order to place the star near the center of the pixel) for all subsequent visits.  The 2012 3.6~\micron~eclipse was observed with a single peak-up at the start of the observation while the 2014 observations used the current standard approach, which includes both an initial peak-up of the position of the star on the array and a second peak-up thirty minutes into the observations in order to correct the initial pointing drift of the tele0scope.  WASP-6 was observed in full array mode with no peak-up pointing, as this mode was not available when these data were taken.  WASP-10 was observed in full array mode with no peak-up pointing in the initial $2010-2011$ observations, while the 2014 observations were obtained in subarray mode with the standard dual peak-up pointing adjustments.  WASP-39 was observed in subarray mode with no peak-up pointing, and WASP-67b was observed in subarray mode with the standard dual peak-up pointing method.  Additional observation details can be found in Table \ref{table:tab2}.

For each data set, we calculate BJD$_{\rm UTC}$ mid-exposure times using information in the BCD files provided by the \textit{Spitzer} pipeline.  We estimate and subtract the sky background, calculate the flux-weighted centroid position of the star on the array, and calculate the total flux in a circular aperture for each individual image as described in previous studies \citep{knutson2012,lewis2013orbital,todorov2013warm,kammer2014search}.  We show the resulting light curves in Fig. \ref{rawfig} and \ref{rawfig2}.  We consider both fixed and time varying photometric aperture sizes in our fits.  For the fixed apertures we consider radii between $2.0-5.0$ pixels, where we step in increments of 0.1 pixels between $2.0-3.0$ pixels and in $0.5$ pixel increments for larger radii.  The time varying apertures were calculated based on a scaling of the noise pixel parameter \citep{mighell2005stellar}, which is proportional to the square of the full width half max of the stellar point spread function, and described by Eq. \ref{eq:npix} below:

\begin{equation}
\beta=\frac{(\sum \limits_{n}I_{n})^{2}}{\sum \limits_{n}I_{n}^{2}}\label{eq:npix}
\end{equation}

where $I_n$ is the measured intensity of the $n^{th}$ pixel. We iteratively re-scale the noise pixel aperture radii as $r = a\sqrt{\beta}+C$, where $a$ is a scaling factor with a value of [0.6, 0.7, 0.8, 0.85, 0.9, 0.95, 1.0, 1.05, 1.1, 1.15, 1.2], and $C$ is a constant with a value between $-0.8$ and $+0.4$ stepping in $0.1$ pixel increments.  

\begin{deluxetable*}{lccccccc}
\tabletypesize{\scriptsize}
\tablecaption{Best Fit Eclipse Parameters\label{table:tab3}}
\tablecolumns{5}
\tablewidth{0pt}
\tablehead{
\colhead{Target} & \colhead{Band ($\mu m$)} & \colhead{$F_p/F_*$ (\%)} & \colhead{$F_p/F_{*,avg}$ (\%)\tablenotemark{a}} & \colhead{T$_{\rm bright}$ (K)\tablenotemark{a}} & \colhead{$T_s$\tablenotemark{b}} & $O-C$ (d)\tablenotemark{c} &\colhead{\textit{e} cos $\omega$} }
\startdata
HAT-P-19b & 3.6 & $0.071\pm0.020$ & $0.062\%\pm0.014\%$ & $1090^{+61}_{-69}$ & $5999.5293\pm0.0074$ &  \phantom{$-$}$0.0062\pm0.0075$ & $\phantom{-}0.0015\pm0.0021$\phantom{$c$}\\
 & 3.6 & $0.053\pm0.021$ & & & $6933.5697\pm0.0071$ &  \phantom{$-$}$0.0013\pm0.0076$ &\\ 
 & 4.5 & $0.066\pm0.019$ & $0.062\%\pm0.016\%$& $914^{+62}_{-71}$ & $5815.1239$\tablenotemark{d}\phantom{$pm0.000$} & &\\ 
 & 4.5 & $0.053\pm0.028$ & & & $6929.5614$\tablenotemark{d}\phantom{$pm0.000$} & &\\
WASP-6b & 3.6 & $0.094\pm0.019$ & &  $1235^{+70}_{-77}$  &  $5196.3606\pm0.0106$ & $-0.0110\pm0.0106$ & \phantom{$-$}$0.0001\pm0.0020$\\
 & 4.5 & $0.115\pm0.022$& &  $1118^{+68}_{-74}$  & $5219.9007\pm0.0044$ &  \phantom{$-$}$0.0021\pm0.0044$ & \\
WASP-10b & 3.6 & $0.103\pm0.015$ & $0.100\%\pm0.011\%$ & $1152^{+34}_{-36}$ & $5580.9226\pm0.0025$ & $-0.1095\pm0.0025$ & $-0.0552\pm0.0007$\phantom{$c$}\\
 & 3.6 & $0.097\pm0.017$ & & & $6910.7917\pm0.0027$ &  $-0.1141\pm0.0027$ &\\ 
 & 4.5 & $0.126\pm0.025$ & $0.146\%\pm0.016\%$ & $1086^{+38}_{-39}$ & $5457.2249\pm0.0040$ &  $-0.0980\pm0.0040$ & \\ 
 & 4.5 & $0.158\pm0.020$ & & & $6923.1693\pm0.0020$ &  $-0.1077\pm0.0020$ &\\
WASP-39b & 3.6 & $0.088\pm0.015$ & & $1213^{+58}_{-62}$  & $6022.2356\pm0.0066$ &  \phantom{$-$}$0.0033\pm0.0068$ & $\phantom{-}0.0007\pm0.0017$\phantom{$c$}\\
 & 4.5 & $0.096\pm0.018$ & & $1055^{+60}_{-65}$ &  $6026.2878\pm0.0058$ &  \phantom{$-$}$0.0005\pm0.0060$ & \\
WASP-67b & 3.6 & $0.014\pm0.018$ & $0.022\%\pm0.013\%$ & $<1046$\tablenotemark{e} & $6860.3124$\tablenotemark{d}\phantom{$pm0.000$} & &\\ 
& 3.6 & $0.032\pm0.020$ & & & $6837.2416$\tablenotemark{d}\phantom{$pm0.000$} & &\\
& 4.5 & $0.088\pm0.024$ & $0.080\%\pm0.018\%$ & $1015^{+67}_{-74}$ & $6846.4689\pm0.0035$ &  \phantom{$-$}$0.0026\pm0.0041$ & $\phantom{-}0.0007\pm0.0012$\phantom{$c$}\\
& 4.5 & $0.070\pm0.026$ & & & $6869.5390\pm0.0064$ &  \phantom{$-$}$0.0007\pm0.0068$ &
\enddata
\tablenotetext{a}{We report the error-weighted mean eclipse depths at 3.6 and 4.5~\micron~for planets with multiple observations in each bandpass.  We note that the averaged 3.6~\micron~eclipse depth for WASP-67b is formally a non-detection with a significance of $1.7\sigma$.  Brightness temperatures are calculated using a PHOENIX stellar model interpolated to match the published temperature and surface gravity for each star.}
\tablenotetext{b}{BJD$_{\rm UTC}$ - 2,450,000.  To convert from UTC to TDB time standards, simply add 66.2~s to eclipse times obtained between $2009-2012$, and 67.2~s to eclipse times from 2014.}
\tablenotetext{c}{Observed minus calculated eclipse times, where we have accounted for the uncertainties in both the measured and predicted eclipse times as well as the added delay from the light travel time in the system.  This delay is 61~s, 54~s, 49~s, 63~s, and 66~s for HAT-P-19b, WASP-6b, WASP-10b, WASP-39b, and WASP-67b, respectively.}
\tablenotetext{d}{We allow the eclipse times in this bandpass to vary as free parameters in our fit, but we use the error-weighted mean orbital phase and corresponding uncertainty from the other bandpass to place a prior constraint on their values.  This reduces the uncertainty in the measured eclipse depths for visits where the eclipse depth is not detected at a statistically significant level ($>3\sigma$; see \S\ref{sec:eclipsemodel} for more details).}
\tablenotetext{e}{$2\sigma$ upper limit based on the error-weighted average of the two 3.6~\micron~eclipse measurements.}
\end{deluxetable*}

We account for variations in intrapixel sensitivity using a pixel-level decorrelation (PLD) method, which \cite{deming2014spitzer} found to produce results that are superior to those of a simple polynomial decorrelation or pixel mapping method.  Details of this PLD technique are described in Deming et al., but the basic methodology is as follows.  First, we obtain the raw flux values for a $3\times3$ grid of pixels centered on the position of the star, and then normalize these individual pixel values by dividing by the total flux in each $3\times3$ postage stamp. This normalization effectively removes any astrophysical signals, and we expect that the remaining variations in the fluxes of individual pixels as a function of time are primarily due to changes in the position of the star on the array. Equation \ref{eq:PLD} shows the resulting forward model, which we fit simultaneously with the eclipse model:

\begin{equation}
F_{model}(t) = \frac{\sum\limits_i w_i F_i (t)}{\sum\limits_i F_i (t)} \label{eq:PLD} 
\end{equation}
where $F_{model}$ is the predicted stellar flux in an individual image, $F_i$ is the measured flux in the $i^{th}$ individual pixel, and $w_i$ is the weight associated with that pixel.  We leave these weights as free parameters in our fit, and solve for the values that best match our observed light curves.  

We consider versions of the photometry with varying bin sizes, aperture sizes, and with up to two hours of data trimmed from the start of the time series in our fits.  By fitting the binned light curves we are able to identify solutions with significantly less red noise on longer time scales and slightly higher scatter on the shortest time scales.  The eclipses of the planets in this study typically have durations of several hours, and we are therefore most sensitive to noise on these time scales in our fits.  For each fit we determine the best-fit model solution using the binned photometry, and then apply our best-fit model to the unbinned light curves in order to generate the plots shown here.  We calculate our best-fit eclipse model and corresponding measurement uncertainties using the version of the photometry that minimizes the amount of time-correlated (``red'') noise in the residuals after the best-fit model has been removed.  As discussed in \cite{deming2014spitzer}, we evaluate the noise properties of a given set of residuals by calculating the root mean variance of the residuals as a function of bin size, where we step through bin sizes in intervals of powers of two.  If the noise is Gaussian and there is no correlation between adjacent data points, we would expect the RMS of the binned data to decrease as $1/\sqrt n$ where $n$ is the number of points in each bin.  We evaluate the relative amount of red noise in each version of the photometry by taking the difference between a model with the ideal $1/\sqrt n$ scaling and the observed RMS at each bin size, squaring the difference, and summing over all bins in order.  By minimizing this least-squares metric, we can select the version of the photometry which has the least amount of red noise across all time scales (see Fig. \ref{rootn_fig} and \ref{rootn_fig2}).   

When considering the optimal choice of aperture and bin size, we first identify the version of the photometry with the lowest overall RMS variance in the best-fit residuals and discard all of the alternative versions of the photometry with a RMS variance that is a factor of 1.2 greater than this value.  This ensures that we avoid solutions that minimize the red noise component by significantly increasing the amount of white noise (these are usually the largest apertures, where the sky background contributes additional noise).  We then pick the bin size and photometric aperture that has the smallest amount of red noise as measured by our least squares metric.  

We find that there is an exponential ramp visible at the start of some visits, which is a well-known feature of the IRAC 3.6 and 4.5~\micron~arrays and has a shape that varies depending on the illumination history of the array immediately prior to each observation as well as the brightness of the target star \citep[e.g.,][]{lewis2013orbital,zellem14}.  We remove this ramp by trimming data from the start of each light curve.  After identifying the optimal bin size and aperture choice, we determine the correct amount of data to trim by examining the best light curve after detector effects have been removed and trimming until no ramp is visible at the start of the observations.  We then revisit our choice of bin size and aperture with the new trim duration in order to make sure that our previous choice is still the best one with the new trim interval.  As shown in Fig. \ref{rootn_fig} and \ref{rootn_fig2}, this results in light curves with no detectible red noise component on the time scales spanned by our observations.  We also calculate a Lomb-Scargle periodogram for the best-fit residuals from our nominal light curve for each visit and find that for periods between five minutes and five hours there are no peaks in the power spectrum with false alarm probabilities lower than $10\%$.  As a final step, we check to make sure that we obtain eclipse depths that are consistent at the $1\sigma$ level across different versions of the photometry that meet our $20\%$ excess noise threshold and have minimal red noise components.

\begin{figure}[h]
\epsscale{1.2}
\plotone{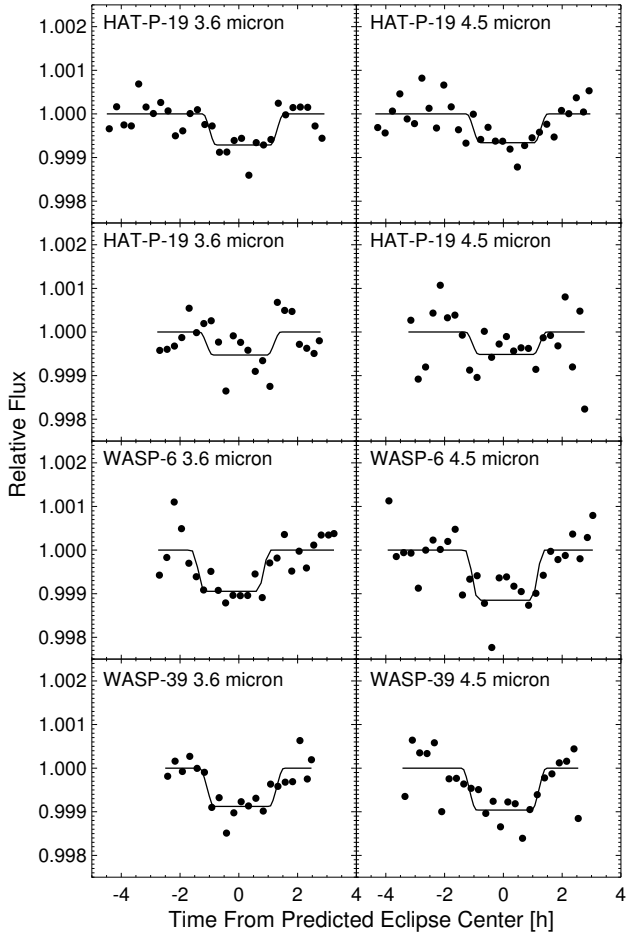}
\caption{Normalized \textit{Spitzer} 3.6 and 4.5 $\mu$m light curves for HAT-P-19, WASP-6, and WASP-39 as a function of time from the predicted center of eclipse, where we have divided out the best-fit instrumental model shown in Fig. \ref{rawfig}. The normalized flux is binned in 10 minute intervals, and best fit eclipse model light curves are over plotted for comparison (solid lines).  Observations for HAT-P-19 are plotted in chronological order, with visits from 2011-2012 one row up from the bottom and visits from 2014 in the bottom row.}
\label{normfig}
\end{figure}

\begin{figure}[h]
\epsscale{1.2}
\plotone{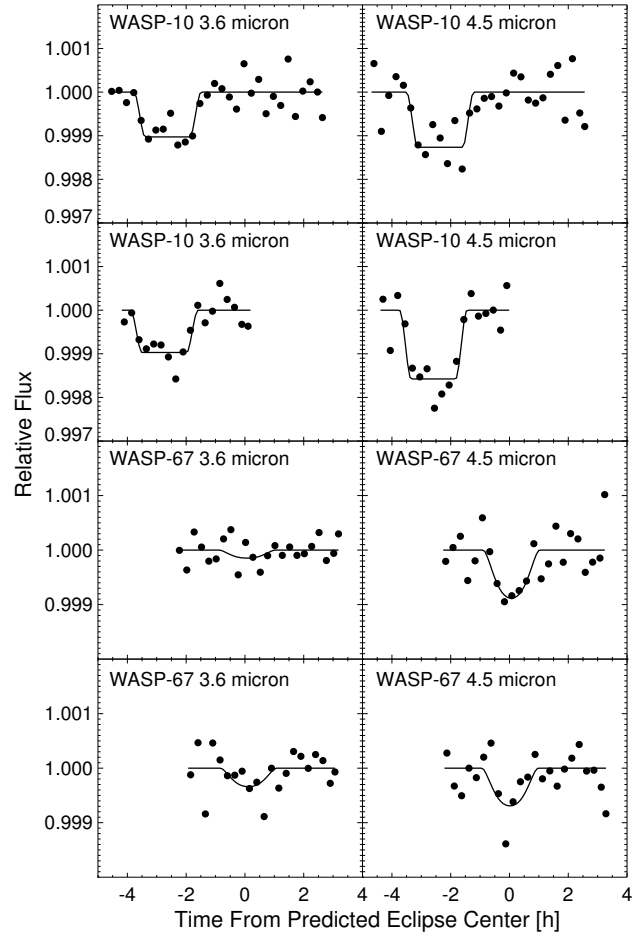}
\caption{Normalized \textit{Spitzer} 3.6 and 4.5 $\mu$m light curves for WASP-10 and WASP-67 as a function of time from the predicted center of eclipse, where we have divided out the best-fit instrumental model shown in Fig. \ref{rawfig2}.  Observations for both planets are plotted in chronological order, with earlier visits in the upper row and later visits in the lower row.  See Fig. \ref{normfig} for more details.}
\label{normfig2}
\end{figure}

\begin{figure*}[ht]
\epsscale{1.15}
\plotone{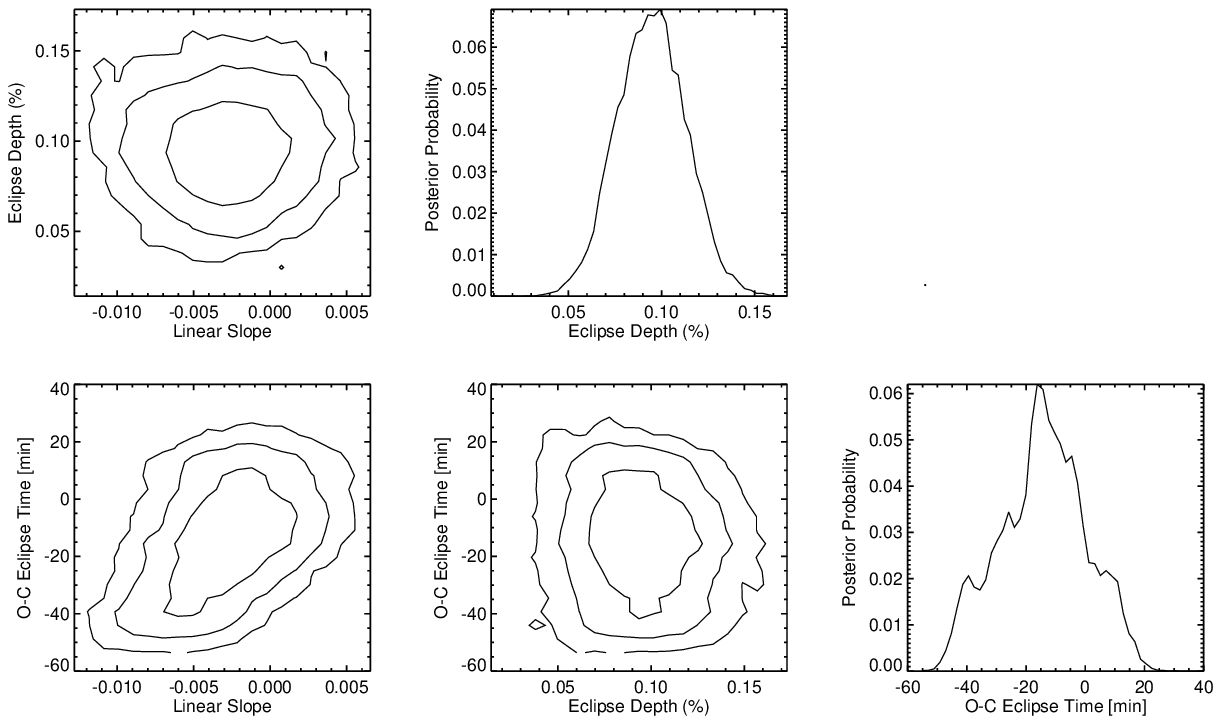}
\caption{Posterior probability distributions and covariances with 1, 2, and 3$\sigma$ contour levels for the secondary eclipse depth, observed minus calculated center of eclipse time, and linear function of time from from our MCMC analysis of the WASP-6b 3.6~\micron~observation.  We calculate the O-C eclipse time assuming that the planet has a circular orbit and using the transit ephemeris from \citet{tregloan15}.  We find no correlation between the measured eclipse depth and the other two parameters, but there does appear to be a correlation between the center of eclipse time and the linear trend slope, which is likely responsible for the larger than usual measurement errors on our reported center of eclipse time.}
\label{wasp6_pdf}
\end{figure*}

\subsection{Eclipse Model and Uncertainty Estimates}\label{sec:eclipsemodel}

We use the routines described in \citet{mandel2002analytic} in order to generate our eclipse light curves, where we set the planet-star radius ratio, orbital inclination, and the ratio of the orbital semi-major axis to the stellar radius equal to their best-fit values from previous studies (see Table \ref{table:tab1} for a complete list of references for each planet).  We assume a circular orbit for the purpose of calculating the eclipse shape and duration, as there is no evidence for an orbital eccentricity large enough to appreciably alter the eclipse duration (\textit{e} sin $\omega>0.1$, where $e$ is the orbital eccentricity and $\omega$ is the longitude of periastron) in any of these systems.  We allow the individual eclipse depths and center of eclipse times to vary as free parameters in our fits.  For the second $4.5$~\micron~eclipse observations of HAT-P-19b and both 3.6~\micron~observations of WASP-67b the eclipse is detected with marginal significance and we therefore place a Gaussian prior on the phase of the secondary eclipse in order to reduce the corresponding uncertainties on the eclipse depth.  We do not fix the center of eclipse time in these fits, but instead implement this prior as a penalty in $\chi^2$ proportional to the deviation from the error-weighted mean center-of-eclipse phase and corresponding uncertainty from the two 3.6 (HAT-P-19b) or 4.5 (WASP-67b)~\micron~eclipses.  

We find that the use of this prior changes the best-fit eclipse depth for the second 4.5~\micron~observation of HAT-P-19b by less than $0.1\sigma$, while reducing the measurement error on this depth by $40\%$ as compared to a fit with no constraint on the center of eclipse phase.  For consistency, we apply the same prior to the first 4.5~\micron~eclipse observation of HAT-P-19b, which is detected with a significance of $3.2\sigma$ in fits with the standard uniform prior on the eclipse phase.  We find that in this fit the use of the prior has no effect on the best-fit eclipse depth and reduces the measurement error in this parameter by $5\%$, as expected for a case where the data themselves provide good constraints on the eclipse time.  For WASP-67b our 3.6~\micron~fits with a uniform prior prefer an eclipse solution at a much earlier phase with a best-fit eclipse depth that is consistent with zero at the $1\sigma$ level.  We therefore conclude that this is likely a local minimum in $\chi^2$ caused by a trade-off with the instrumental noise model, and place a prior on the eclipse phase based on the measured 4.5~\micron~eclipse phase.  This allows us to obtain new constraints on the eclipse depth at 3.6~\micron~for a range of phases consistent with the measured 4.5~\micron~eclipses.

We determine the best fit model for each individual eclipse using a Levenberg-Marquardt minimization routine, where each fit has thirteen free parameters.  In addition to the nine pixel weight parameters and two eclipse parameters, we also fit for a linear function of time in order to account for long-term instrumental and stellar trends.  Figures \ref{rawfig} and \ref{rawfig2} show the raw photometry, while Figures \ref{normfig} and \ref{normfig2} show the corresponding normalized photometry after dividing out the best-fit PLD model and linear function of time.  We estimate the corresponding uncertainties on our model parameters using a Markov chain Monte Carlo (MCMC) analysis with $10^6$ steps per chain, where we define the $1\sigma$ uncertainties as the symmetric interval around the median parameter value containing 68\% of the total probability.  We assume uniform priors on all parameters with the exception of the 3.6~\micron~eclipse phase of WASP-67b and the 4.5~\micron~eclipse phase for HAT-P-19b as noted in the previous paragraph.  

We initialize our chains using the best-fit solution from the Levenberg-Marquardt minimization in order to minimize the burn-in time.  We then check to see where the $\chi^2$ value of the chain first drops below the median value over the entire chain, and trim all points prior to this step in order to remove any residual burn-in phase.  We also examine our resulting posterior probability distributions in order to make sure that our choice of prior bounds is broad enough that it does not affect our final results, and re-run our chain with a broader prior range if required.  Although we find that individual pixel weights are often degenerate with each other, our probability distributions for the best-fit eclipse depths and center of eclipse times are Gaussian and uncorrelated with the other fit parameters, with the exception of the 3.6~\micron~eclipse of WASP-6b which we discuss in more detail below.  The correlation between pixel weights could be mitigated by a transformation to an independent set of basis vectors \citep[e.g.,][]{morello14,morello15}, but as these correlations do not affect our final best-fit eclipse parameters and their corresponding uncertainties we do not pursue this option here.

We find that our measured center of eclipse time for WASP-6b's 3.6~\micron~eclipse has an uncertainty that is significantly larger (15 minutes) than is typical for our other observations with statistically significant eclipse detections.  We examined the results from our MCMC analysis and identified a correlation between the slope of the linear trend across the 3.6~micron~eclipse and the measured center of eclipse time in this visit (see Fig. \ref{wasp6_pdf}).  As discussed earlier, we found it necessary to trim the first 1.5 hours of data in order to remove an unusually long-lived ramp at the start of these observations.  This trimming means that we have a relatively short baseline prior to the start of the eclipse, and it is therefore not surprising that there might be a correlation between the eclipse time and the linear trend in our fits.  Our best-fit eclipse depth for this visit does not appear to be correlated with either of these parameters, and we find no evidence for a similar correlation between eclipse time and linear trend slope in other visits with short pre-eclipse baselines.

\section{Discussion}\label{sec:xDiscussion}

\begin{figure}[h]
\epsscale{1.1}
\plotone{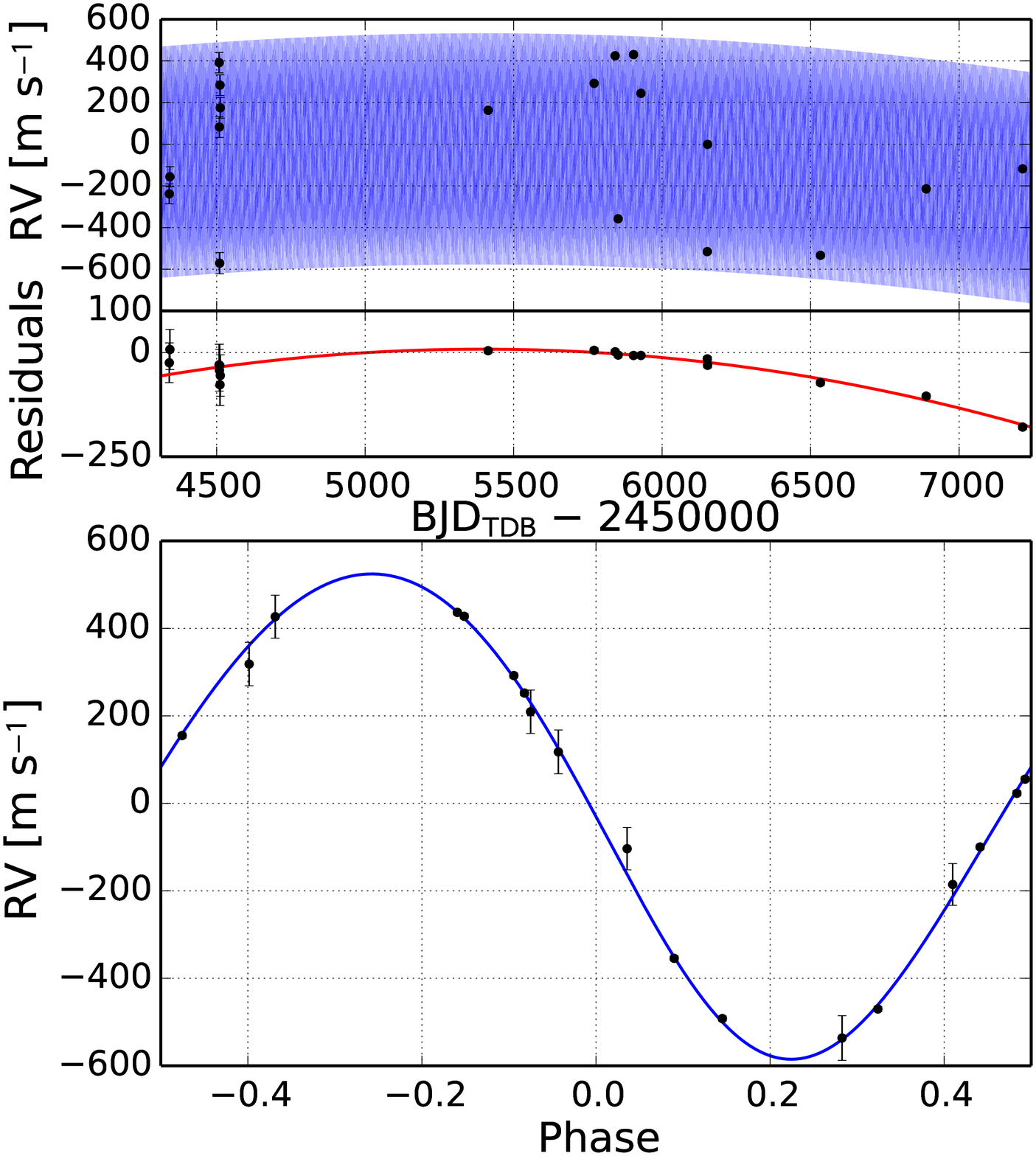}
\caption{The top panel shows radial velocity measurements (filled black circles) for WASP-10 from \citet{christian09} and \citet{knutson2014b}, with best-fit model over plotted in blue.  The residuals after subtracting the orbit of the transiting hot Jupiter WASP-10b are shown in the middle panel, with best-fit radial velocity trend overplotted in red.  Although a linear trend was reported in \citet{knutson2014b}, the addition of two new Keck HIRES radial velocity measurements required us to fit a curved trend in this analysis.  The best-fit orbital solution for WASP-10b is shown in blue in the lower panel, with phased radial velocity measurements after removal of the best-fit trend over plotted as filled black circles.  See \citet{knutson2014b} for additional details.}
\label{rv_plot}
\end{figure}

\subsection{Orbital Eccentricity}\label{ecc_disc}
WASP-10b is the only planet in our study that was previously known to have an eccentric orbit, with $e=0.0473\pm0.0032$ based on a total of sixteen radial velocity measurements from Keck HIRES and SOPHIE \citep{christian09,knutson2014b}.  Our error-weighted average secondary eclipse phase of $0.4649\pm0.004$ confirms this planet's non-zero orbital eccentricity, and we combine our measurement with published radial velocity data from \citet{knutson2014b} in order to obtain improved constraints on the orbital eccentricity $e$ and longitude of periastron $\omega$ (see Fig. \ref{rv_plot}).  We also add two new Keck HIRES measurements that extend the baseline of these data by approximately a year and a half beyond the last measurement reported in our previous paper.  These new measurements clearly establish the presence of curvature in the observed radial velocity trend, and we have adapted our fits accordingly.

We give the updated orbital parameters from a joint fit to the radial velocity measurements and using the measured transit ephemeris and secondary eclipse times in Table \ref{wasp10_rv_table} \citep[see][for additional information on the fitting method]{knutson2014b}.  Our new orbital eccentricity of $e=0.0589^{+0.0033}_{-0.0026}$ differs from our previous value by $4\sigma$; this change can be attributed to the inclusion of our measured secondary eclipse times in the new fits as well as the switch to a curved radial velocity trend model.  We note that a previous study reported a detection of the secondary eclipse for this planet in the $K$s band at an orbital phase of $0.4972\pm0.0005$ \citep{cruz15}; this offset differs significantly from that of our \textit{Spitzer} secondary eclipse observations and is also inconsistent with the current radial velocity measurements for this system.  The authors of this study note that their reported error is likely an underestimate as it does not account for additional uncertainties contributed by the removal of systematic noise sources from their light curve.  See \S\ref{fortney_models} for additional discussion of this observation.

\begin{deluxetable}{lrr}
\tablecaption{Updated Orbital Solution for WASP-10 from Radial Velocity Fit\label{wasp10_rv_table}}
\tablehead{\colhead{Parameter} & \colhead{Value} & \colhead{Units}}
\startdata
$P_{b}$ & 3.0927278 $\pm 2.8e-06$ & days\\
$T_{\textrm{conj},b}$ & 2454664.03809 $\pm 4.7e-05$ & \bjdtdb\\
$K_{b}$ & 556.7 $^{+5.4}_{-5.6}$ & \ms\\
$e\cos{\omega}$ & -0.054 $\pm 0.001$ & \\
$e\sin{\omega}$ & 0.027 $^{+0.0071}_{-0.007}$ & \\
$e_{b}$ & 0.0608 $^{+0.0036}_{-0.003}$ & \\
$\omega_{b}$ & 153.6 $^{+6.2}_{-5.7}$ & degrees\\
$\gamma_{1}$ & 15.3 $\pm 2.7$ & \ms\\
$\gamma_{2}$ & -64 $\pm 41$ & \ms\\
$\dot{\gamma}$ & -0.037 $\pm 0.013$ & \ms day$^{-1}$\\
$\ddot{\gamma}$ & -5.94e-05 $^{+1e-05}_{-9.9e-06}$ & \ms day$^{-2}$\\
jitter & 6.3 $^{+1.7}_{-1.3}$ & \ms\\
\enddata
\tablenotetext{}{These fits used the transit ephemeris from \citet{barros13} and the secondary eclipse times reported in this paper as priors on the fit in order to obtain improved constraints on the orbital solution.  Radial velocity data taken from \citet{christian09} and \citet{knutson2014b}, with the addition of two new unpublished RV measurements from Keck/HIRES.  We allow the zero points of each radial velocity data set ($\gamma_1$ and $\gamma_2$ to vary independently in our fits, and describe the radial velocity trend with a linear ($\dot{\gamma}$) and quadratic ($\ddot{\gamma}$) function of time.  The reference epoch for $\gamma$,$\dot{\gamma}$,$\ddot{\gamma}$ is 2455615.0}
\end{deluxetable}

Published radial velocity observations for the remaining four planets in this study suggest that their orbits are likely circular, in good agreement with our measured secondary eclipse times.  WASP-6b has the most extensive radial velocity data set, with a total of 79 measurements of which 38 were obtained in a single night in order to measure the Rossiter-McLaughlin effect as the planet passed in front of its host star \citep{gillon2009discovery}.  Although the initial discovery paper reported a non-zero orbital eccentricity of $0.054\pm0.017$, this finding was subsequently disputed by \citet{husnoo2012observational}.  These authors reanalyzed the same data while excluding the single night corresponding to the Rossiter measurement and found a best-fit orbital eccentricity of $0.04\pm0.02$.  For HAT-P-19b the radial velocity data constrain the orbital eccentricity to a value of $0.07\pm0.04$, and for WASP-39b the orbital eccentricity was fixed to zero in the published fits \citep{faedi2011wasp}.   \textit{Spitzer} secondary eclipse observations provide a significantly tighter constraint on the value of $e \cos \omega$, where $e$ is the orbital eccentricity and $\omega$ is the longitude of periastron.  These eclipse times can be used to further refine the global orbital eccentricity when fitting radial velocity data \citep[e.g.,][]{knutson2014b}.  We find that the observed secondary eclipse times for these four planets are consistent with circular orbits, and list the corresponding $e \cos \omega$ values in Table \ref{table:tab3}.

\begin{figure}[ht]
\epsscale{1.1}
\plotone{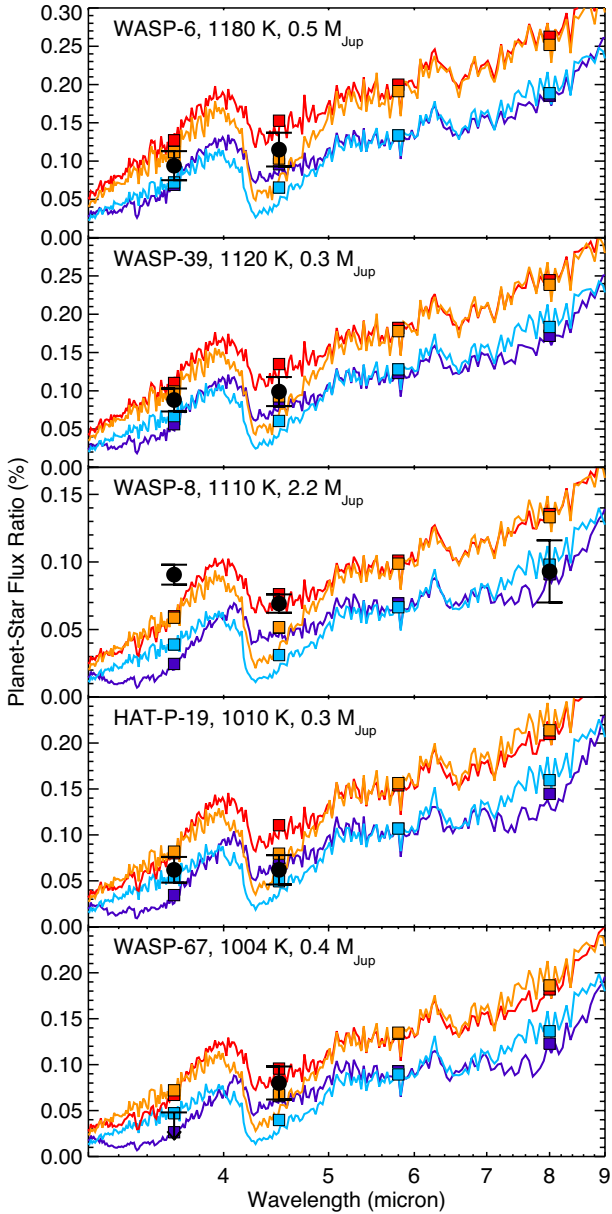}
\caption{Predicted planet-star flux ratio as a function of wavelength for 1x solar (purple) and 1000x solar (light blue) atmospheric metallicities and efficient recirculation of energy from the day side to the night side \citep{fortney2008unified}.  We also show the corresponding 1x solar (red) and 1000x solar (orange) models for inefficient recirculation.  Both models assume a dayside albedo of zero, where a non-zero albedo would have the same effect as increasing the efficiency of the day-night circulation.  Models are shown as solid lines, with the corresponding band-integrated model predictions overplotted as filled squares.  Measured \textit{Spitzer} secondary eclipse depths with their associated 1$\sigma$ uncertainties are shown as black filled circles, with the exception of the 3.6~\micron~eclipse of WASP-67b where we plot the $2\sigma$ upper limit.  In addition to the three planets described in this paper, we also compare these same models to published secondary eclipse depths for GJ~436b \citep{lanotte2014}, HAT-P-20b \citep{deming2014spitzer}, WASP-8b \citep{cubillos2013wasp,deming2014spitzer}, and WASP-80b \citep{triaud15}.  We list masses and predicted equilibrium temperatures for each planet at the time of secondary eclipse calculated assuming zero albedo and efficient redistribution of energy to the night side.}
\label{ratiofig}
\end{figure}

\begin{figure}[ht]
\epsscale{1.1}
\plotone{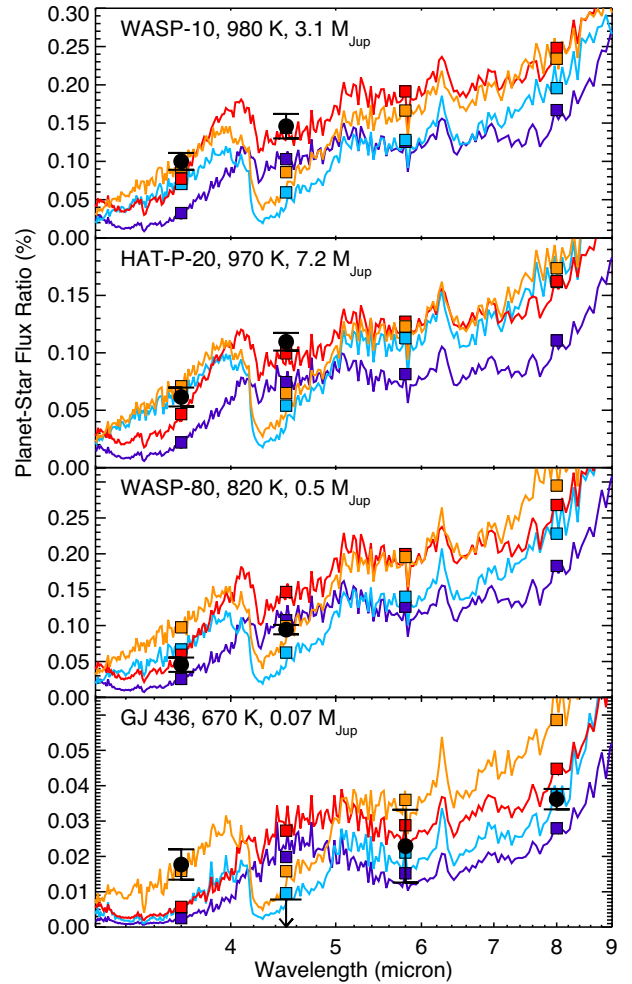}
\caption{Predicted planet-star flux ratio as a function of wavelength for 1x solar and 1000x solar atmospheric metallicities and either efficient or inefficient recirculation of energy from the day side to the night side (see Fig. \ref{ratiofig} for more information).  Measured \textit{Spitzer} secondary eclipse depths with their associated 1$\sigma$ uncertainties are shown as black filled circles, with the exception of the 4.5~\micron~eclipse of GJ 436bb where we plot the $2\sigma$ upper limit.  }
\label{ratiofig2}
\end{figure}

\subsection{Comparison to Atmosphere Models}\label{model_disc}

We next compare the measured \textit{Spitzer} secondary eclipse depths for each planet to the predictions of standard one dimensional atmosphere models. We consider two classes of models, based on \cite{burrows2008theoretical} and \cite{fortney2008unified}. Both models assume that the chemistry of these atmospheres is in local thermodynamic equilibrium and parameterize the unknown recirculation of energy to the night side.  

\subsubsection{Constraints on Atmospheric Metallicity and Day-Night Circulation}\label{fortney_models}

We use the \cite{fortney2008unified} models to explore a range of atmospheric metallicities from $1-1000\times$ solar.  For planets at these temperatures, an increase in the atmospheric metallicity results in a decrease in the relative amount of methane as compared to CO and CO$_2$.  Because methane absorbs in the 3.6~\micron~\textit{Spitzer} bandpass while CO and CO$_2$ absorb in the 4.5~\micron~bandpass, metal-rich models will have correspondingly strong emission at 3.6~\micron~and weak emission at 4.5~\micron~as compared to their solar metallicity counterparts.  At $1\times$ solar, the pressure-temperature profiles in the models presented here are in radiative equilibrium and are self-consistent between the predicted equilibrium chemistry mixing ratios, the opacity of these molecules, and the the absorption and emission of flux.  At $1000\times$ solar, the models use the radiative-equilibrium pressure-temperature profiles from a $50\times$ solar model where we have recalculated the chemical equilibrium and spectra at $1000\times$ solar metallicity.  Although not fully self-consistent, these models are a reasonable approximation for the appearance of a very metal-enhanced atmosphere.    We test the validity of these models in the high metallicity regime by comparing them to the measured secondary eclipse depths for GJ~436b \citep{lanotte2014}.  This planet appears to be best-matched by a model with 1000x solar metallicity in good agreement with previous studies \citep[e.g.,][]{moses2013compositional}, but we are not able to fully match the steep slope across the 3.6 and 4.5~\micron~bands.  We expect that models including disequillibrum chemical process and/or non-solar carbon to oxygen ratios \citep[e.g.,][]{visscher11,moses11} could provide an improved match to these observations, and leave this as an exercise for future modeling studies.

Figure \ref{ratiofig} and \ref{ratiofig2} compare our measured secondary eclipse depths for the five planets in this study to 1x and 1000x solar metallicity models with either efficient or inefficient redistribution of energy to the night side.  The relative efficiency of recirculation is implemented by varying the amount of flux incident at the top of the one-dimensional atmospheric column, and is therefore equivalent to a change in the planet's dayside albedo. For the inefficient case, the incident flux is set equal to the dayside average value, while for the efficient case the incident flux is calculated as the average over the entire surface.  For four of the five new planets, our observations are best-matched by models with efficient recirculation of energy to the night side but we are unable to distinguish between low and high metallicity atmosphere models with high statistical significance.  WASP-10b appears to be the exception to this trend, and is best matched by models with inefficient recirculation and a relatively low atmospheric metallicity.  It is also by far the most massive planet in our sample, a fact that we discuss in more detail below.  Although we do not show it in these plots, there is a published $K$s (2.1~\micron) secondary eclipse measurement available for this planet with a value of $0.137\%\pm0.016\%$ \citep{cruz15}.  The measured planet-star flux ratio in this bandpass is $3-10\times$ larger than the predictions of our models, and the reported center of eclipse phase differs from our measurement by 2.4 hours ($54\sigma$), comparable to the eclipse duration.  It is therefore unlikely that this ground-based measurement corresponds to the same eclipse signal detected in our \emph{Spitzer} photometry, and in fact the authors note that their reported uncertainties are likely underestimated due to the unknown contribution of systematic noise sources to their error budget.

Given that we obtain relatively tight constraints on the atmospheric properties of WASP-10b, it is interesting to ask whether or not it is representative of other Jovian planets in this temperature range.  A search of the literature indicates that there are two other planets (HAT-P-20b and WASP-8b) with measured \textit{Spitzer} secondary eclipse depths at 3.6 and 4.5~\micron, masses greater than that of Jupiter, and predicted equilibrium temperatures less than 1200 K.  We compare the observations for these two planets to the same set of models as before, and plot the results in Figure \ref{ratiofig} and \ref{ratiofig2}.  For HAT-P-20b the data are best matched by the solar metallicity model with inefficient circulation of energy to the night side, consistent with the conclusions from  \citet{deming2014spitzer} and \citet{triaud15}.  For WASP-8b \citep{cubillos2013wasp,deming2014spitzer} we also prefer models with inefficient energy recirculation and solar metallicity, but this model still under-predicts the measured flux in the 3.6~\micron~band.  The unusual nature of WASP-8b's dayside emission spectrum is discussed in detail in \citet{cubillos2013wasp}, and may be related to its eccentric orbit ($e=0.30$).  The only other eccentric planets in this sample are GJ 436b and HAT-P-20b, which have orbital eccentricities of 0.15 and 0.016, respectively  \citep{knutson2014b}.  Despite this discrepancy, the data for these three planets suggest a clear pattern: planets with masses greater than that of Jupiter are best described by models with solar metallicity and relatively inefficient day-night circulation.

We next consider whether or not the smaller planets in our sample show a different behavior.  As noted earlier, HAT-P-19b, WASP-6b, WASP-39b, and WASP-67b are all best-matched by models with relatively efficient day-night circulation and are consistent with a range of atmospheric metallicities.  We identify two additional planets in the literature with measured \textit{Spitzer} secondary eclipse depths at 3.6 and 4.5~\micron, masses greater than that of Jupiter, and predicted equilibrium temperatures less than 1200 K.  As before, we compare the observations for these planets to our standard set of models, and plot the results in Figure \ref{ratiofig} and \ref{ratiofig2}.  For WASP-80b \citep{triaud15}, we find the data require a model with relatively efficient day-night circulation and an atmosphere with at most a moderately enhanced metallicity.  For GJ 436b we obtain the best match using a model with efficient atmospheric circulation (this may be related to the planet's non-zero orbital eccentricity) and a high atmospheric metallicity as discussed above.

\begin{figure}[ht]
\epsscale{1.1}
\plotone{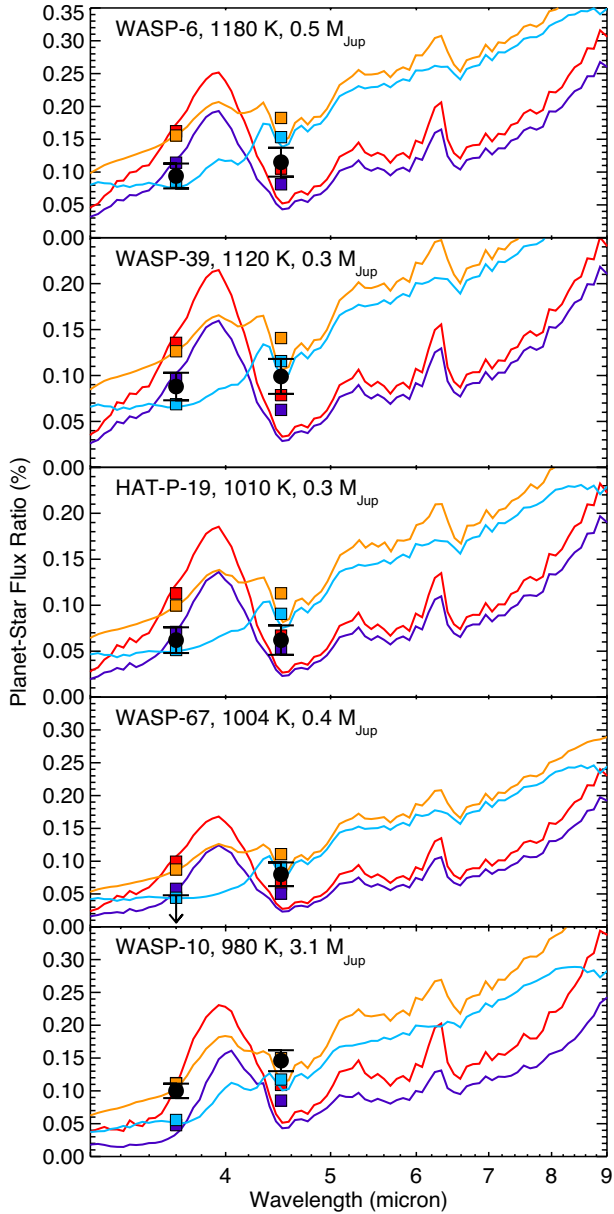}
\caption{Solar metallicity atmosphere models showing the effect of a temperature inversion \citep{burrows2008theoretical}.  Model with efficient day-night circulation ($P_n=0.1$) and either a dayside temperature inversion ($\kappa=0.1$ cm$^2$ g$^{-1}$; light blue) or no inversion ($\kappa=0.0$ cm$^2$ g$^{-1}$; purple) are shown as solid lines with band-integrated values over plotted as filled squares.  All models assume an albedo of zero for the planet's dayside.  We also plot the corresponding models with inefficient day-night circulation ($P_n=0.4$) and either a dayside temperature inversion (orange) or no inversion (red).  Measured \textit{Spitzer} secondary eclipse depths with their associated 1$\sigma$ uncertainties are shown as black filled circles, with the exception of the 3.6~\micron~eclipse of WASP-67b where we plot the $2\sigma$ upper limit.  }
\label{burrowsratio}
\end{figure}

\subsubsection{Temperature Inversions, Clouds, and Other Caveats}

Although the models described above suggest the presence of some intriguing patterns, it is important to note that with measurements in just two bandpasses we cannot provide unique constraints on the atmospheric chemistries of these planets.  In addition to the potential effects of disequilibrium chemistry and non-solar C/O ratios, the vertical thermal profiles of these planets can also alter the shapes of their dayside emission spectra.  We explore the potential effects of temperature inversions using a suite of models from \citet{burrows2008theoretical}, which allow us to artificially create atmospheric temperature inversions with varying strengths.  Temperature inversions are created by adding a generalized gray absorber in the stratosphere above 0.03 bars with an absorption coefficient $\kappa$ of 0.1 cm$^2$ g$^{-1}$.  A dimensionless parameter $P_n$ describes the efficiency of energy redistribution, with $P_n = 0.0$ indicating recirculation on the dayside only, and $P_n = 0.5$ indicating energy uniformly distributed between the two hemispheres.  In these models the redistribution is implemented via a heat sink located at pressures between $0.1-0.01$ bars.  

As shown in Fig. \ref{burrowsratio}, our measured eclipse depths for for four of the five planets presented in this paper require models with relatively efficient day-night circulation but we are unable to distinguish between models with and without a temperature inversion at a statistically significant level.  As with the previous set of models, WASP-10b is best described by models with relatively inefficient day-night circulation.  We note that the solar metallicity models with no inversion, which should be identical to the Fortney solar metallicity models, in fact display significantly stronger molecular absorption bands.  This difference has been consistently observed in models for planets across a wide range of temperatures \citep[e.g.,][]{deming11,todorov12,todorov2013warm,orourke14,shporer14}, and may be due to either the use of different opacity tables, a change in their pressure-temperature profiles due to differences in their method for incorporating heat redistribution to the planet's night side, or some combination of the two.

We also consider the possibility that the dayside emission spectra of these planets may be affected by the presence of high altitude clouds or hazes, which have been observed in the transmission spectra of several cool gas giant planets \citep{line13,jordan13,knutson2014a,nikolov15}.  A range of condensate clouds are possible at these temperatures and pressures \citep[for a recent review see][]{marley13}, some of which may occur high enough up in the atmosphere to have a detectible effect.  If the atmospheres of these planets have significant methane, then it may also be possible to form a layer of hydrocarbon soot via photochemistry in the upper layers of the atmosphere \citep{kempton12,morley13}.  Published optical transmission spectra for WASP-6b indicate that it does indeed possess a high altitude cloud layer \citep{jordan13,nikolov15}, but there are no published transmission spectra available for HAT-P-19b, WASP-10b, WASP-39b, or WASP-67b.  If these planets possess a cloud layer that is optically thick when viewed during secondary eclipse, we would expect such a cloud layer to suppress the strengths of the observed molecular absorption or emission features, leading to a more blackbody-like spectrum.  HAT-P-19b has a dayside emission spectrum that is moderately inconsistent with that of a pure blackbody, which should produce a planet-star flux ratio that increases consistently towards longer wavelengths.  WASP-6b, WASP-10b, WASP-39b, and WASP-67b are closer to a blackbody shape, and we therefore cannot exclude the possibility that their dayside emission spectra are affected by clouds.  However, we note that the transit technique is sensitive to more tenuous clouds and hazes than the secondary eclipse technique, as the slant optical path length traveled by light transmitted through the planet's atmosphere is longer than the radial path traveled by light emitted by the planet \citep{fortney2005}.  GJ 436b serves as a useful case in point, as it has an apparently featureless infrared transmission spectrum \citep[e.g.,][]{knutson2014a} but still exhibits strong spectral features in its dayside emission spectrum \citep[e.g.,][]{stevenson2010possible,lanotte2014}.  

\begin{figure}[ht]
\epsscale{1.2}
\plotone{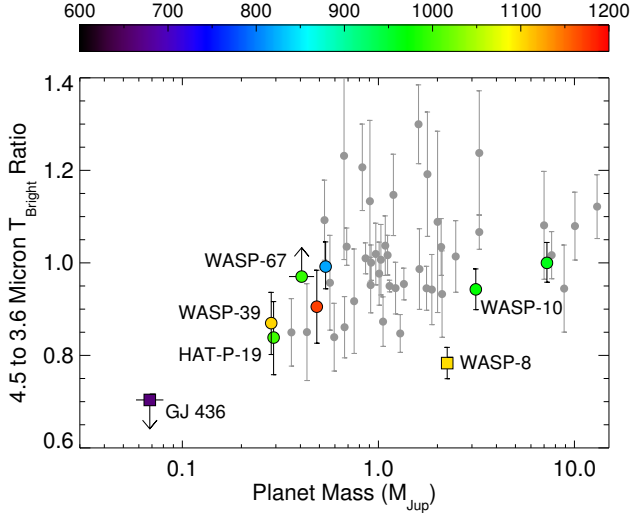}
\caption{Ratio of the 4.5 to 3.6~\micron~brightness temperatures as a function of planet mass for the nine planets with predicted equilibrium temperatures cooler than 1200 K (colored points) or hotter than 1200 K (grey filled circles) and eclipse depths available in both \emph{Spitzer} bands.  Brightness temperature ratios are shown as filled circles (circular orbits) or squares (orbital eccentricities greater than 0.1; $e=0.30$ for WASP-8b and $e=0.15$ for GJ~436b) where the color corresponds to the predicted equilibrium temperature of each planet at the time of secondary eclipse for the case of zero albedo and efficient redistribution of energy to the night side.  Temperatures for the cool planets are taken from this study, and values for the hot planets are taken from \citet{hansen14}.  The 3.6~\micron~eclipse of WASP-67b and the 4.5~\micron~eclipse of GJ 436b are not detected at a statistically significant level, and we therefore plot the corresponding $2\sigma$ limits on the brightness temperature ratios for these planets.  We note a general trend of increasing temperature ratio as planet mass increases; this trend appears to flatten out for planets more massive than Jupiter, consistent with the prediction that planets in this mass range should have envelopes that reflect the local metallicity of the protoplanetary disk.  WASP-8b is an outlier in this plot, which may be related to its relatively large orbital eccentricity.}
\label{empirical_plot}
\end{figure} 

\subsection{Empirical Comparisons to the Sample of Observed Exoplanet and Brown Dwarf Emission Spectra}

Although our constraints on the atmospheric properties of planets in individual cases are limited by the signal to noise of our measurements, it is possible that the aggregate sample may display trends that are not apparent on a case-by-case basis.  We therefore examine the properties of the planets in our sample using a series of model-independent metrics that allow us to construct a view of the global properties of this population.  We begin by calculating the ratio of the 4.5 to 3.6~\micron~brightness temperatures for each planet; this ratio is independent of both the planet-star radius ratio and the effective temperatures of the planet and star, and instead simply reflects the relative shape of each planet's emission spectrum as determined by the presence of any molecular absorption or emission features.  If we assume that molecular features are observed in absorption (i.e., none of the planets have dayside temperature inversions), we would expect the ratio of the 4.5 to 3.6~\micron~brightness temperatures to vary primarily as a function of atmospheric metallicity.  In this scenario metal-rich planets will have relatively small brightness temperature ratios due to strong absorption from CO at 4.5~\micron~and weak absorption from methane at 3.6~\micron, while metal-poor planets will have correspondingly large ratios due to strong methane absorption and weak CO absorption.  

\begin{figure}[ht]
\epsscale{1.2}
\plotone{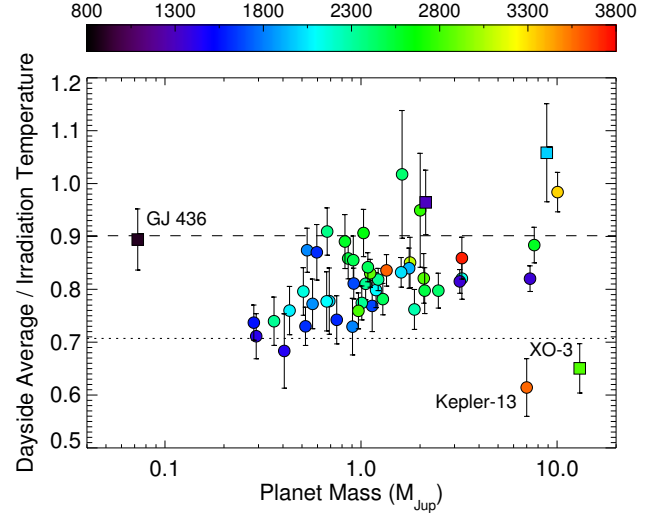}
\caption{Ratio of the effective dayside temperature derived from measured secondary eclipse depths to the irradiation temperature (predicted temperature at the substellar point for the case of zero albedo and no heat recirculation) as a function of planet mass for 51 transiting planets with published infrared secondary eclipse depths including the five new planets in this study.  Planets with orbital eccentricities less than 0.1 are shown as filled circles where the color indicates their irradiation temperature, while planets with eccentricities greater than 0.1 are shown as filled squares.  The dayside and irradiation temperatures of these eccentric planets vary as a function of orbital phase, and is therefore not expected to follow a simple scaling relation.  The predicted temperature ratio for the case of no redistribution of energy and zero albedo is shown as a dashed line, and the corresponding ratio for complete redistribution to the night side with zero albedo is shown as a dotted line.}  
\label{temp_ratio_vs_mass}
\end{figure} 

\begin{deluxetable*}{lcccccc}
\tabletypesize{\scriptsize}
\tablecaption{Distance modulus, distance, apparent and absolute magnitudes for HAT-P-19b, WASP-6b, WASP-10b, WASP-39b, and WASP-67b\label{table:CMdata}}
\tablewidth{0pt}
\startdata
  \hline
  \hline
   & Distance Modulus  
& Distance &\multicolumn{2}{c}{Apparent Magnitude}
& \multicolumn{2}{c}{Absolute Magnitude} 
\\
  &  (m-M) &  (pc)
& $m_{\rm [3.6]}$  & $m_{\rm [4.5]}$ &  $M_{\rm [3.6]}$
& $M_{\rm [4.5]}$
\\
  \hline
  \\
HAT-P-19b &$6.73\pm0.25$&
$222\pm28$  &$18.51\pm0.23$ &$18.57\pm0.31$
 &$11.79\pm0.53$  &$11.85\pm0.55$\\
WASP-6b &$6.65\pm0.12$&
$214\pm12$  &$17.84\pm0.22$ &$17.69\pm0.21$
 &$11.19\pm0.38$  &$11.04\pm0.36$\\
WASP-10b &$5.58\pm0.22$&
$131\pm14$  &$17.43\pm0.12$ &$17.09\pm0.12$
 &$11.85\pm0.34$  &$11.50\pm0.33$\\
WASP-39b &$6.67\pm0.12$&
$216\pm12$  &$17.80\pm0.19$ &$17.73\pm0.21$
 &$11.12\pm0.33$  &$11.06\pm0.35$\\
WASP-67b &$6.64\pm0.14$&
$213\pm14$  &$>18.32$ &$17.82\pm0.25$
&$>11.68$ 
&$11.18\pm0.37$
\enddata
\end{deluxetable*}

We plot the measured brightness temperature ratios of the nine planets with published \textit{Spitzer} secondary eclipse depths and effective temperatures cooler than 1200 K as a function of planet mass and equilibrium temperature in Fig. \ref{empirical_plot}.  Interestingly, we see no evidence for a correlation with planet equilibrium temperature, but instead see tentative evidence for an increase in this ratio as planet mass increases from that of Neptune up to Jupiter.  This is consistent with the hypothesis that lower mass planets are more likely to have metal-rich atmospheres, while planets with masses comparable to or larger than that of Jupiter should have envelopes that reflect the (near-solar) metallicity of their local region of the protoplanetary disk.  We obtain a nearly identical plot if we simply take the ratio of the measured eclipse depths in these two bands, indicating that the varying effective temperatures of the planets and their host stars have a relatively small effect on the measured eclipse depths as compared to changes in the observed molecular absorption or emission features.   

We evaluate the statistical significance of this trend by taking the error-weighted average brightness ratio for planets with masses less than 0.3 M$_{\rm Jup}$ and comparing this ratio to the error-weighted average ratio for planets with masses greater than 0.5 M$_{\rm Jup}$.  We exclude GJ 436b and WASP-67b from this calculation, as we only have lower and upper limits on their brightness ratios, respectively.  We also exclude WASP-8 in light of its large orbital eccentricity, which may affect the observed properties of its dayside atmosphere.  We find that planets with masses less than 0.3 M$_{\rm Jup}$ have an average brightness temperature ratio that is 2.1$\sigma$ smaller than the average ratio for the more massive planets in our sample.  We conclude more observations will be required in order to determine whether or not the observed trend is statistically significant.

We next evaluate the relative efficiency of circulation between the day and night sides of these planets by comparing their observed versus predicted dayside emission temperatures as described in \citet{cowan11} and \citet{schwartz15}.  For this exercise we consider both the subset of planets with relatively low irradiation levels including the five planets from this study, and also the full set of fifty transiting gas giant planets with broadband infrared eclipse depths detected with greater than $3\sigma$ significance.  We calculate the irradiation temperature, $T_0 = T_{*}\sqrt{R_*/a}$, for each planet based on published values for the stellar effective temperature and the planet's scaled semi-major axis.  This temperature corresponds to the predicted temperature at the planet's substellar point assuming an albedo of zero and no redistribution of energy.  The maximum hemisphere-averaged dayside effective temperature is then $(2/3)^{1/4}$ of this value, and the temperature corresponding to complete redistribution of energy across the entire planet surface would be $(1/4)^{1/4}$ of this value. 

We calculate the effective dayside temperature measured for each planet using all available infrared ($\lambda > 1~\mu$m) eclipse depth measurements.  We convert the set of individual dayside brightness temperatures across all bandpasses into a combined dayside effective temperature using two different schemes: error-weighted mean (wavelengths with small uncertainties contribute most to the estimate) and power-weighted mean (wavelengths that capture large fractions of the planetary emission contribute most to the estimate).  The uncertainty in the averaged dayside effective temperature and in its ratio with the planet's irradiation temperature is then estimated using a 1000-step Monte Carlo method with equal numbers of trials drawn from each scheme \citep[see][for more details]{schwartz15}.  We use the ratio of the measured vs predicted (irradiation) dayside temperatures to determine whether or not the correlation between planet mass and day-night circulation efficiency inferred in our cool planet sample is consistent with the larger sample of published secondary eclipse data for transiting gas giant planets with a range of incident flux levels.  We plot this ratio as a function of planet mass in Fig. \ref{temp_ratio_vs_mass}, and find that our new observations clearly indicate the existence of a correlation between planet mass and day-night circulation efficiency for planets with lower irradiation levels.  This correlation also appears to be consistent with the published measurements for more highly irradiated planets, although there are fewer low-mass planets in this sample.  We conclude that lower-mass planets may be more reflective and/or better at transporting heat to their night sides.  Although the current data do not allow us to distinguish between these two hypotheses,  the combination of visible-light secondary eclipse observations and infrared phase curve measurements could provide a definitive answer to this question in future studies.

\begin{figure}[ht]
\epsscale{1.2}
\plotone{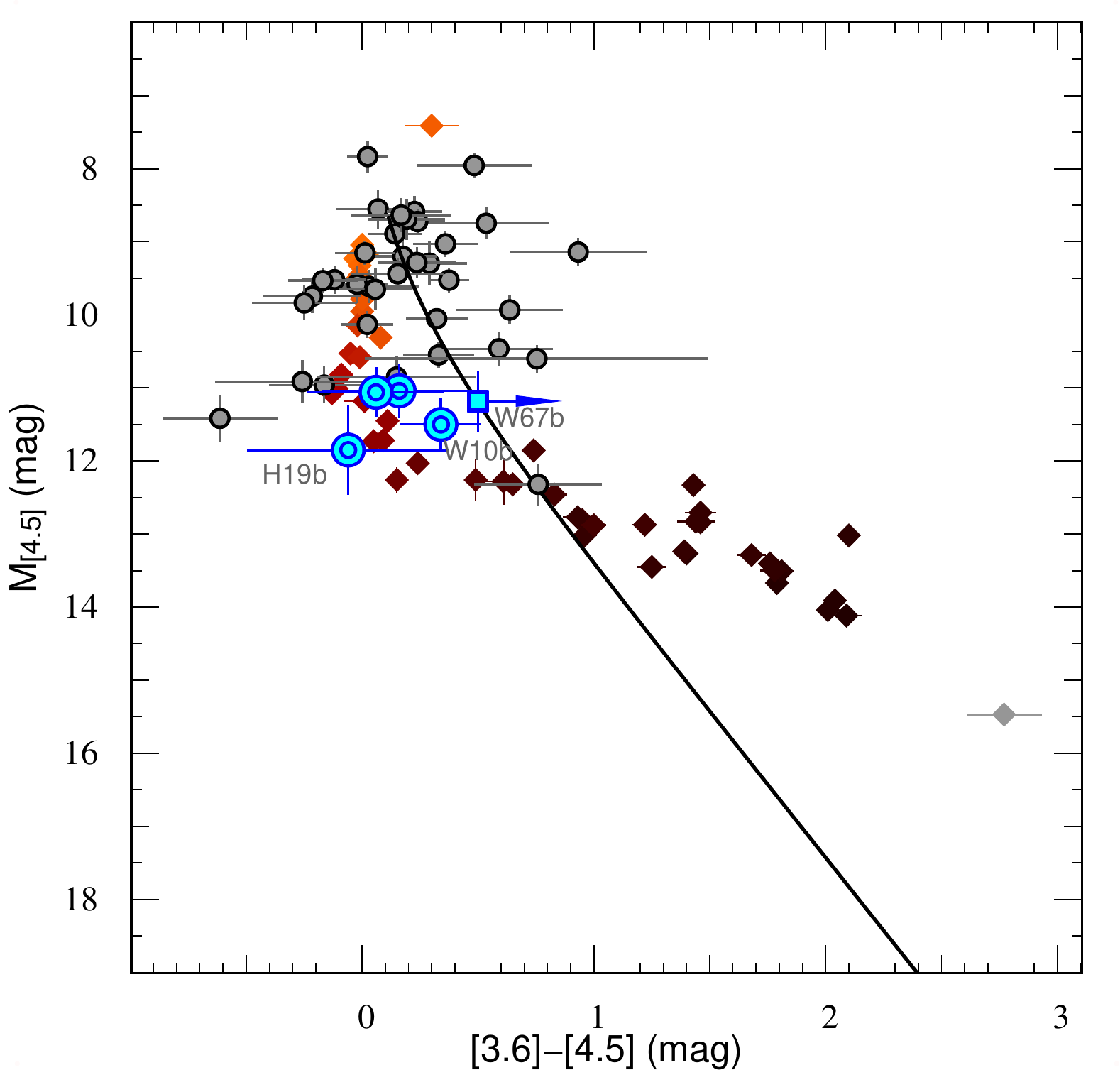}
\caption{Diagram comparing the 4.5~\micron~absolute magnitudes and 3.6$-$4.5~\micron~colors of HAT-P-19b, WASP-6b, WASP-10b, WASP-39b, and WASP-67b (blue filled circles and square) to those of previously published planets and brown dwarfs in the same magnitude range \citep[figure adapted from][]{triaud14b}.  Published planets are plotted as grey circles,  while ultra-cool dwarfs with spectral types between M5$-$Y1 obtained from \citet{dupuy12} are plotted as diamonds with colors indicating their spectral type (orange corresponds to hotter stars and brown indicates cooler stars).  WASP-67b is not detected in the 3.6~\micron~band and is therefore is represented with a blue square and arrow to indicate the corresponding $2\sigma$ lower limit on its 3.6$-$4.5~\micron~color.  The black line indicates the color of a 1 R$_{\rm Jup}$ object emitting as a blackbody over a range of effective temperatures less than 4000 K.}
\label{fig:CM}
\end{figure}

Lastly, we create a color magnitude plot for the five planets in this study that allows us to compare them to ultracool brown dwarfs and directly imaged planets spanning the same range of effective temperatures using the same procedure as described in \citet{triaud14a} and \citet{triaud14b}.  In order to maintain consistency with the other planets in this diagram, we obtain stellar parameters for the five planets presented in this paper from TEPCAT \citep{southworth11} and compute their absolute magnitudes using relations by \citet{flower96} and \citet{torres10}. Distance moduli were calculated by comparing them to visual magnitudes, which we take from the UCAC4/APASS catalog \citep{zacharias13}. Distance moduli, derived distance, visual and absolute magnitudes are provided in Table~\ref{table:CMdata} and visually represented in Figure~\ref{fig:CM}.  We calculate these quantities using WISE 1 and WISE 2 apparent magnitudes \citep{cutri13}, which are a good match for the {\it Spitzer} 3.6 and 4.5~\micron~bands \citep{triaud14b}, and list the resulting values in Table \ref{table:CMdata}.

As shown in Fig. \ref{fig:CM}, all five planets have similar magnitudes in this diagram, albeit with varying colors.  Along with GJ 436b \citep{lanotte2014} and WASP-80b \citep{triaud15}, these five planets have the smallest absolute magnitudes of any extrasolar planets observed to date and demonstrate the benefit of extending the current \textit{Spitzer} observations to smaller and cooler planets.   Three of the five planets appear to be bluer than a blackbody with the same absolute magnitude, placing them near the late L dwarfs in this diagram.  WASP-6b and WASP-67b lie on or near the blackbody line, where the $2\sigma$ upper limit on the 3.6~\micron~eclipse depth for WASP-67b indicates that it is almost certainly redder than a blackbody.  As discussed in \citet{triaud14b}, irradiated exoplanets have optical properties that appear to be more diverse than those of self-luminous brown dwarfs, as demonstrated by the varying colors of the planets in our sample.  Although part of this variation likely reflects intrinsic variations in surface gravities and metallicities \citep[e.g.,][]{zahnle14}, our analysis in \S\ref{model_disc} suggests that disequilibrium chemistry, cloud formation, and variations in atmospheric circulation patterns may also contribute to the observed diversity of dayside emission spectra for the transiting planet sample.

\section{Conclusions}\label{sec:xConclusions}

We present new secondary eclipse depth measurements for HAT-P-19b, WASP-6b, WASP-10b, WASP-39b, and WASP-67b in the 3.6 and 4.5~\micron~\textit{Spitzer} bands.  We use our measured center of eclipse times to constrain the orbital eccentricities of these planets and find that HAT-P-19b, WASP-6b, WASP-39b, and WASP-67b all appear to have circular orbits.  Our new secondary eclipse times confirm the previously reported non-zero orbital eccentricity for WASP-10b, and we carry out a joint fit to all of the available transit, secondary eclipse, and radial velocity data in order to provide improved constraints on this planet's orbit.  

We combine our observations with published secondary eclipse depths for other transiting gas giant planets and find that the most massive planets are best matched by models with low albedos and relatively inefficient day-night circulation, while smaller planets are best matched by models with higher albedos and/or efficient day-night circulation.  Previous analyses of the available secondary eclipse data for transiting hot Jupiters \citep{cowan11,perez13,schwartz15} have found that planets with lower irradiation levels appear on average to have more efficient recirculation of energy than their more highly irradiated counterparts, but did not note any correlation with planet mass.  Our results are consistent with our conclusions in \citet{schwartz15} and \citet{wong15}, where we examined the sample of planets with infrared phase curve observations and found that the two most massive planets in this sample appear to have systematically lower albedos than their less massive counterparts as inferred from a simple energy balance calculation.  In the future, measurements of the secondary eclipses of these planets at optical wavelengths will help to establish whether or not the observed variations are primarily due to changes in the Bond albedos of these planets or to differences in their atmospheric circulation.

All of the planets in this study have predicted equilibrium temperatures cooler than 1200 K, and we therefore expect that the ratio of methane to CO in their atmospheres should vary as a function of their atmospheric metallicity.  When we combine our observations with published secondary eclipse data for four additional planets in this same temperature range, we find that planets with masses greater than that of Jupiter are best matched by relatively low metallicity atmosphere models.  Although our results are less well constrained for the smaller planets in our sample, we examine their aggregate properties by taking the ratio of the 3.6 to 4.5~\micron~brightness temperatures and plotting this ratio as a function of planet mass and temperature.  We find that there is no detectible correlation between this ratio and the equilibrium temperatures of these planets, but we do see tentative evidence for a trend in mass consistent with the existence of an inverse correlation between planet mass and atmospheric metallicity.  We note that with just two wavelengths of data our ability to constrain the atmospheric metallicities of these planets is degenerate with other atmospheric properties, including the presence or absence of dayside temperature inversions as well as the potential effects of high altitude cloud layers.  

Future observations of the dayside emission spectra of these planets spanning a broader range of wavelengths will be able to determine whether or not the observed trends are in fact due to variations in atmospheric metallicity as suggested here.  Although these planets are too cold and faint to have detectible secondary eclipses in the near-infrared wavelengths accessible to the \textit{Hubble Space Telescope} and ground-based telescopes, future spectroscopic observations at longer wavelengths with the \textit{James Webb Space Telescope}  will provide invaluable information about their atmospheric compositions.  For planets without high altitude clouds or hazes, measurements of water absorption in transit with WFC3 on \textit{HST} can also provide independent constraints on the inferred atmospheric metallicity \citep[e.g.,][]{kreidberg2014,benneke15}.  In the short term, observations of additional planets in the \textit{Spitzer} bands will help to determine whether or not the observed correlation between planet mass and 3.6 to 4.5~\micron~brightness temperature ratio is statistically significant, and will allow us to better understand the differences between cool transiting gas giant planets and the sample of directly images planets and ultra-cool brown dwarfs with similar temperatures. 

\acknowledgements{
J.-M.~D. and N.~K.~L. acknowledge funding from NASA through the Sagan Exoplanet Fellowship program administered by the NASA Exoplanet Science Institute (NExScI). This work is based on observations made with the \textit{Spitzer Space Telescope}, which is operated by the Jet Propulsion Laboratory, California Institute of Technology, under contract with NASA.}

\end{document}